\documentclass[runningheads]{llncs}
\usepackage{hyperref}
\usepackage{pgfplots}
\pgfplotsset{compat=newest}
\usepackage{graphicx}
\usepackage{url}
\usepackage[export]{adjustbox}
\usepackage{rotating}
\usepackage{booktabs, makecell, tabularx}
\usepackage{tabularx,ragged2e}

\usepackage[strings]{underscore}
\usepackage{diagbox}
\usepackage{multirow}
\usepackage{setspace}
\usepackage{pdfpages}

\begin{document}
\title{Deep Two-Stage High-Resolution Image Inpainting}
%
%
\author{Andrey Moskalenko \and Mikhail Erofeev \and Dmitriy Vatolin}
\authorrunning{A. Moskalenko et al.}
%
\institute{Lomonosov Moscow State University, Moscow, Russia
\email{\{andrey.moskalenko,merofeev,dmitriy\}@graphics.cs.msu.ru}}
\maketitle              
\begin{abstract}
In recent years, the field of image inpainting has developed rapidly, learning based approaches show impressive results in the task of filling missing parts in an image. But most deep methods are strongly tied to the resolution of the images on which they were trained. A slight resolution increase leads to serious artifacts and unsatisfactory filling quality. These methods are therefore unsuitable for interactive image processing. In this article, we propose a method that solves the problem of inpainting arbitrary-size images. We also describe a way to better restore texture fragments in the filled area. For this, we propose to use information from neighboring pixels by shifting the original image in four directions. Moreover, this approach can work with existing inpainting models, making them almost resolution independent without the need for retraining. We also created a GIMP plugin that implements our technique. The plugin, code, and model weights are available at \url{https://github.com/a-mos/High_Resolution_Image_Inpainting}.

\end{abstract}
\section{Introduction}
Image inpainting is the process of realistically filling unknown or damaged regions of an image. An inpainting algorithm receives as input a corrupted image and a mask; its output is a restored image. 


\begin{figure}[ht]
\centering
\setlength\tabcolsep{1pt}
\settowidth\rotheadsize{Radcliffe Cam}
\setkeys{Gin}{width=\hsize}
\begin{tabularx}{1.0\linewidth}{l XXXXX}
\rothead{\centering \begin{math}{512 \times 512}\end{math}}       
&   \includegraphics[valign=m]{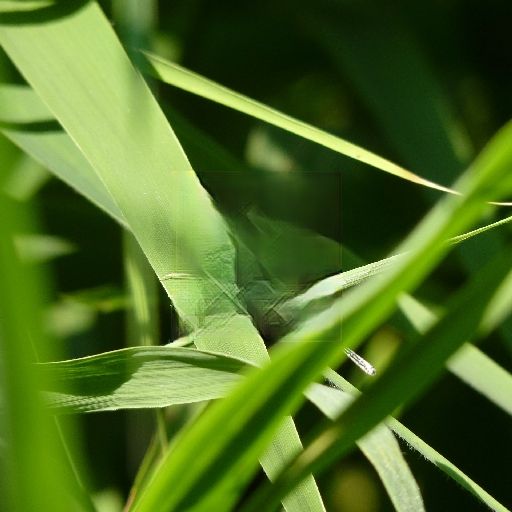}
&   \includegraphics[valign=m]{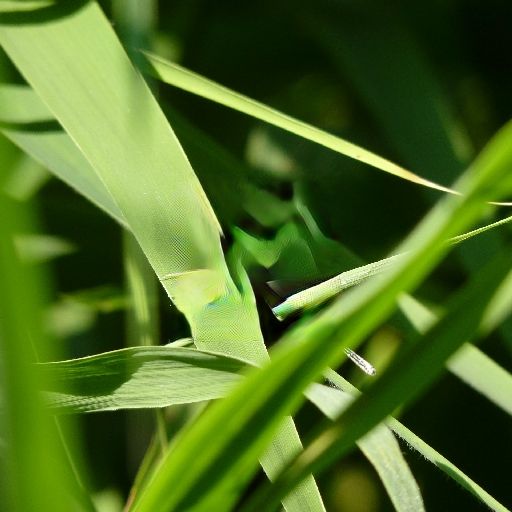}  
&   \includegraphics[valign=m]{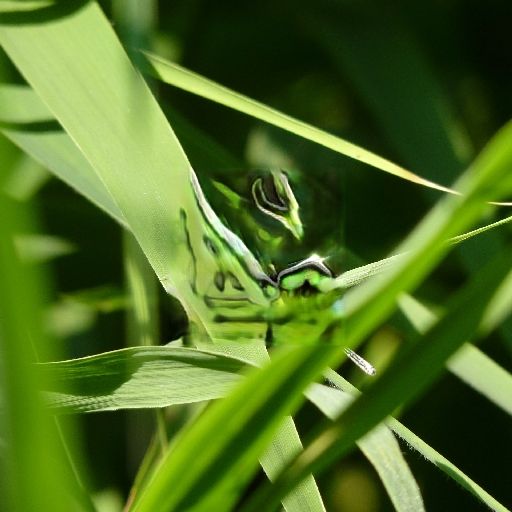}
&   \includegraphics[valign=m]{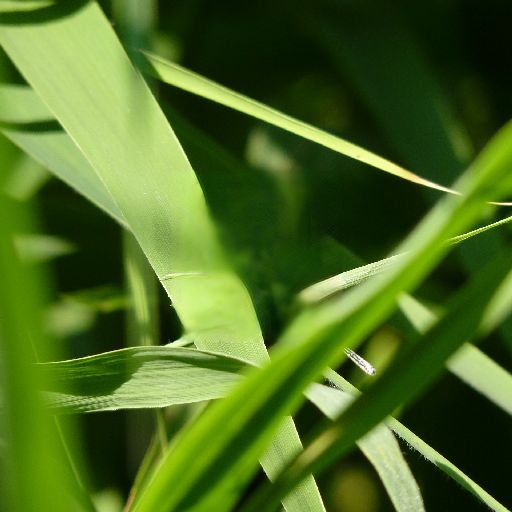} 
&   \includegraphics[valign=m]{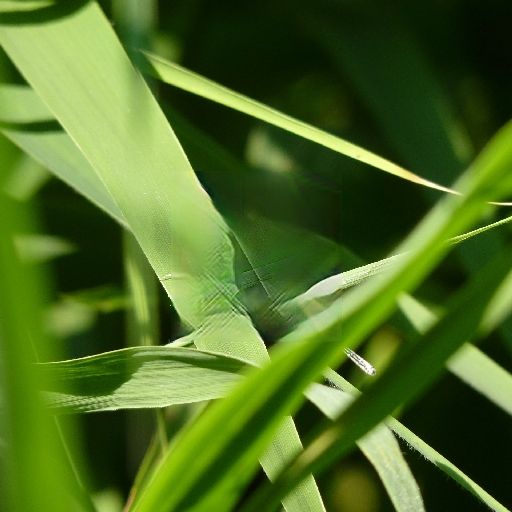} \\
\addlinespace[2pt]
\rothead{\centering \begin{math}{1024 \times 1024}\end{math}} 
&   \includegraphics[valign=m]{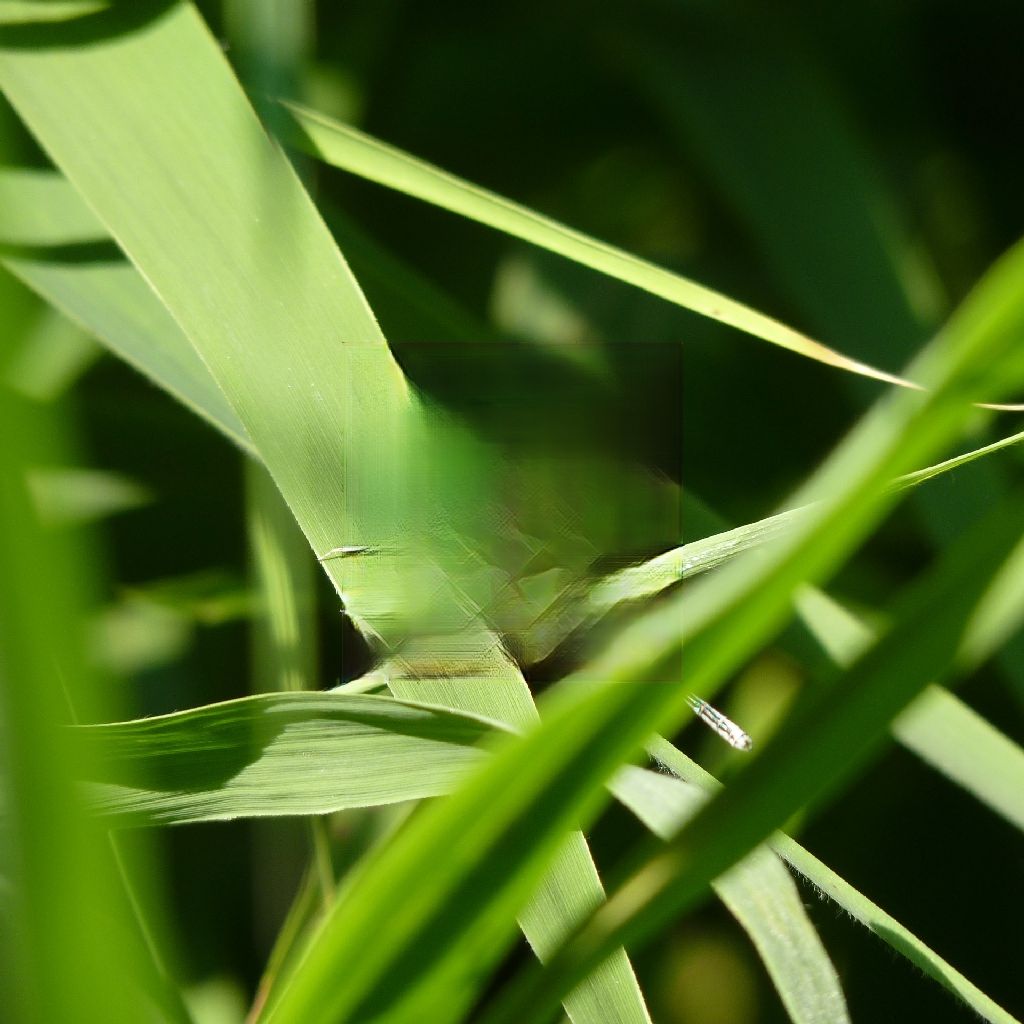}
&   \includegraphics[valign=m]{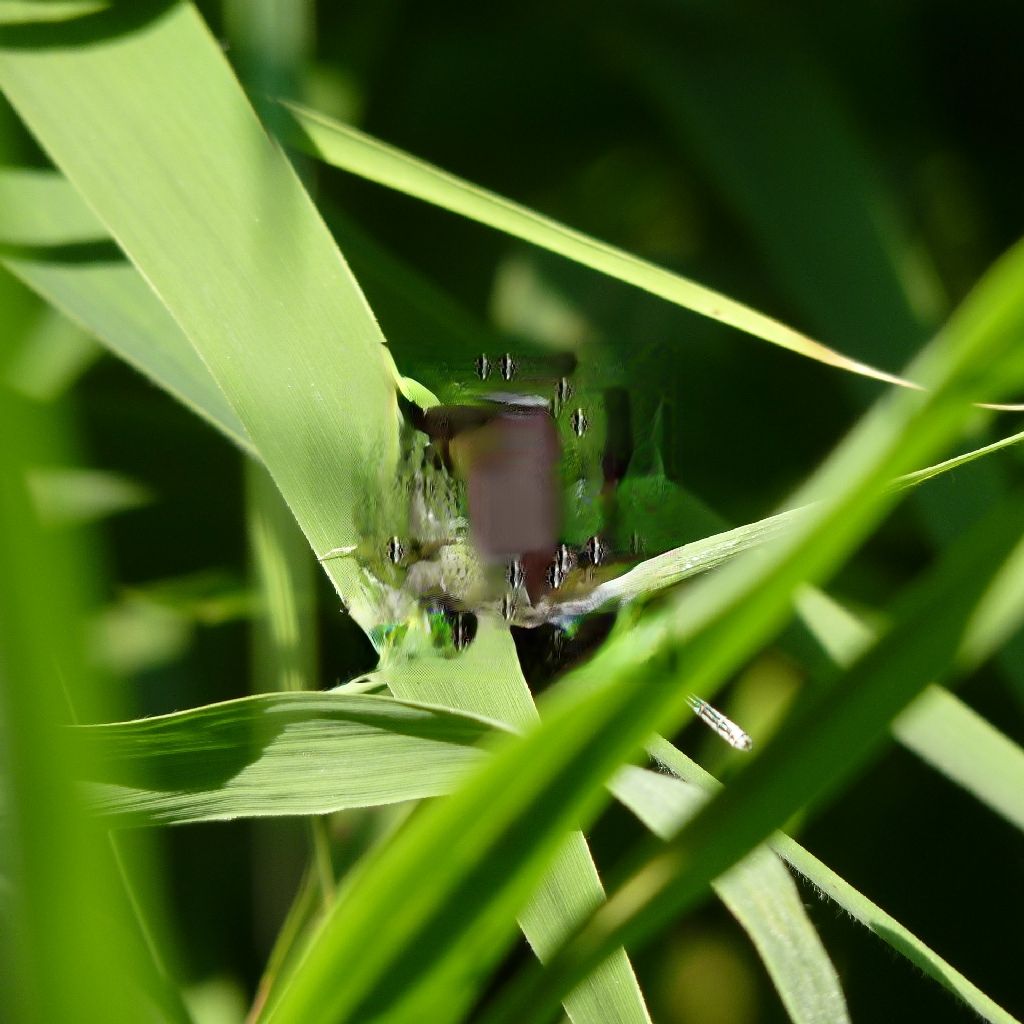}    
&   \includegraphics[valign=m]{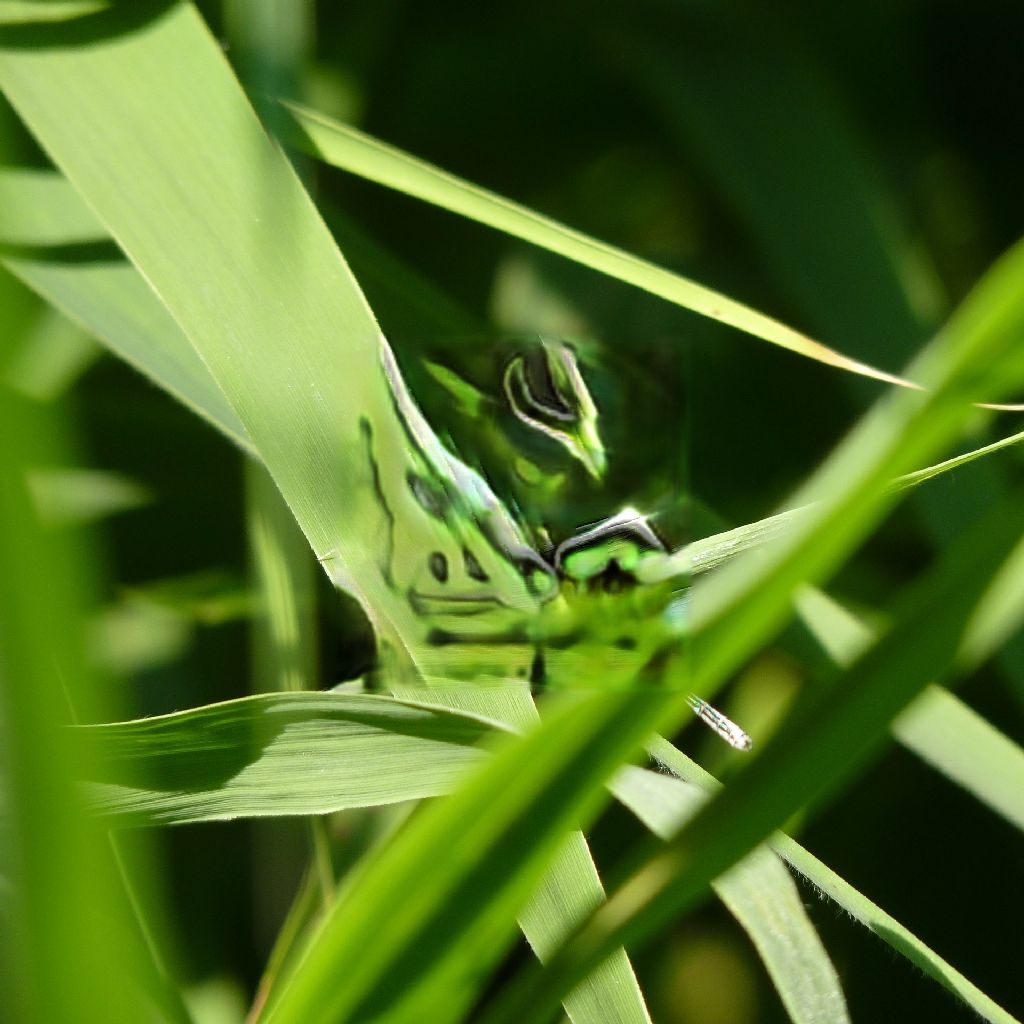}
&   \includegraphics[valign=m]{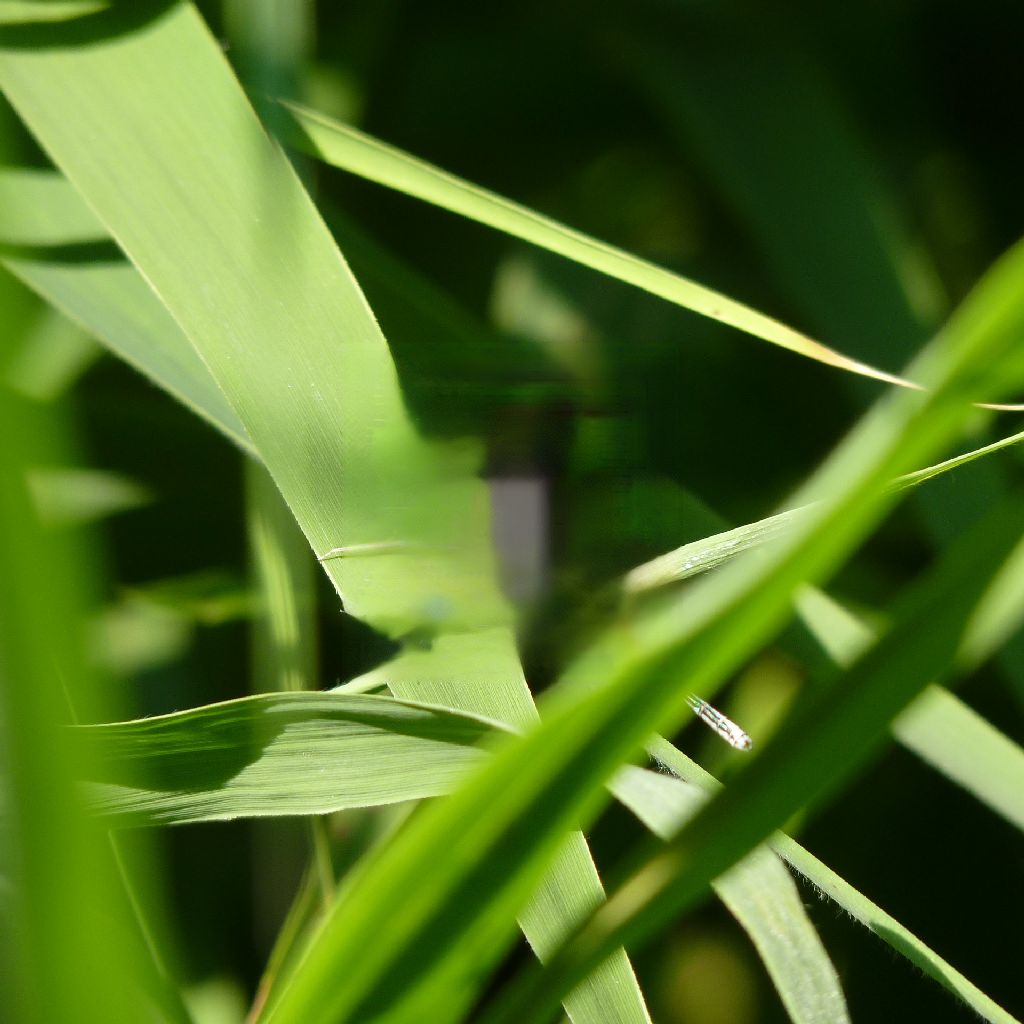}
&   \includegraphics[valign=m]{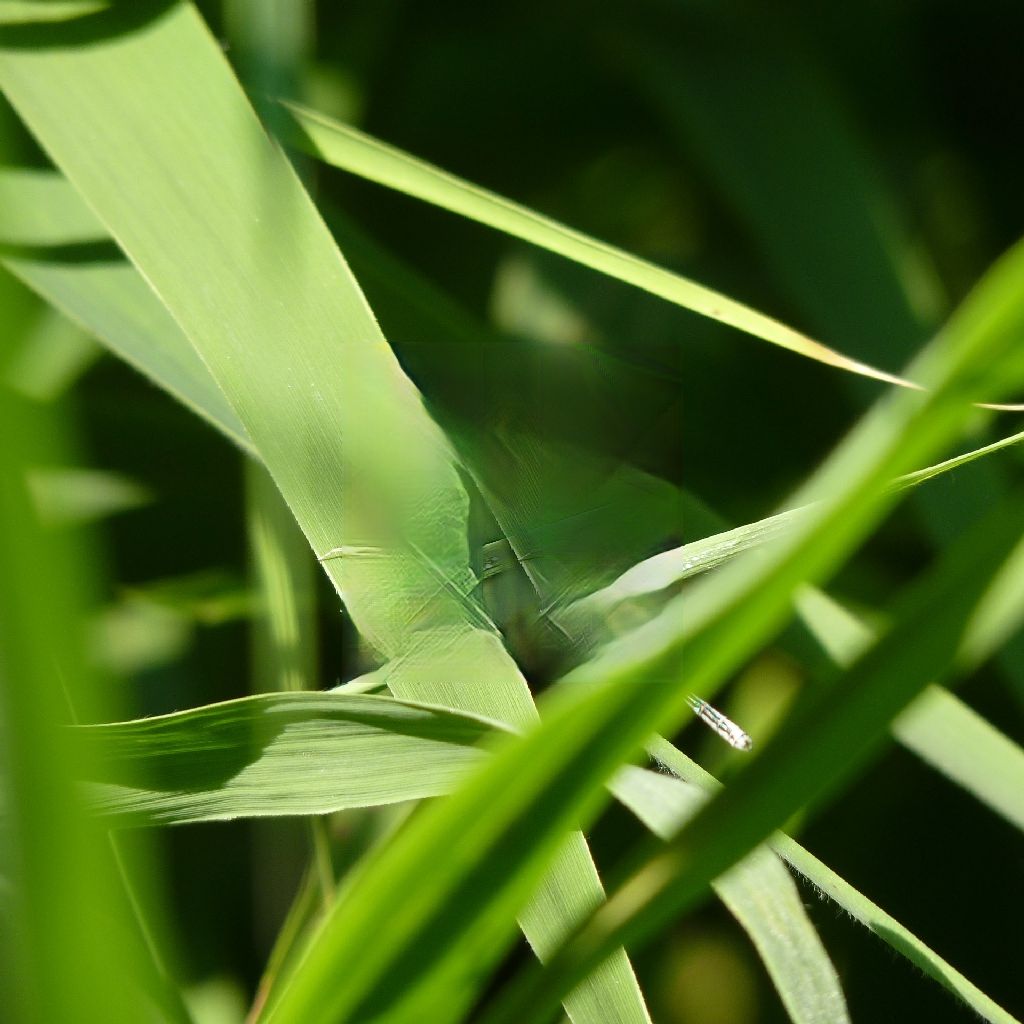}\\

\addlinespace[4pt]
\rothead{\centering \begin{math}{512 \times 512}\end{math}}       
&   \includegraphics[valign=m]{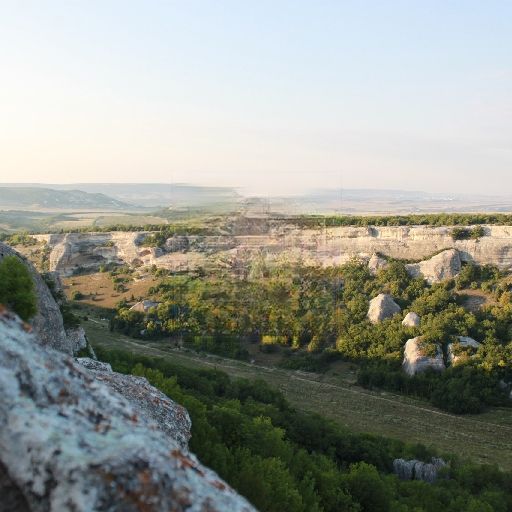}
&   \includegraphics[valign=m]{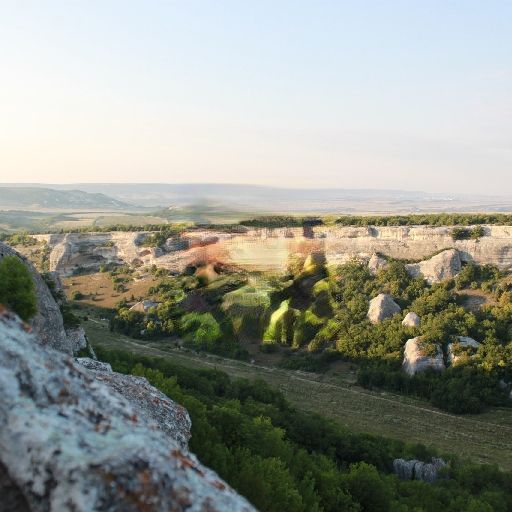}   
&   \includegraphics[valign=m]{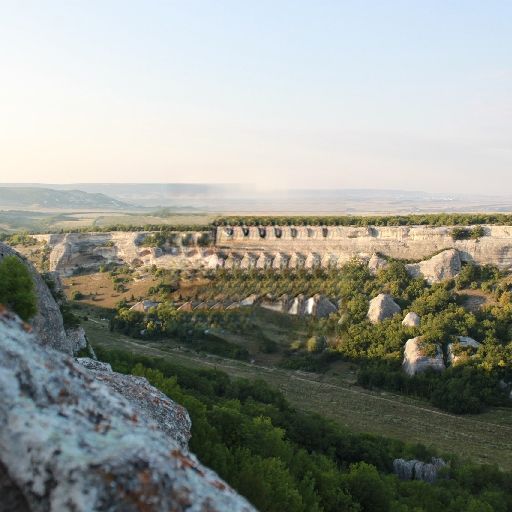} &   \includegraphics[valign=m]{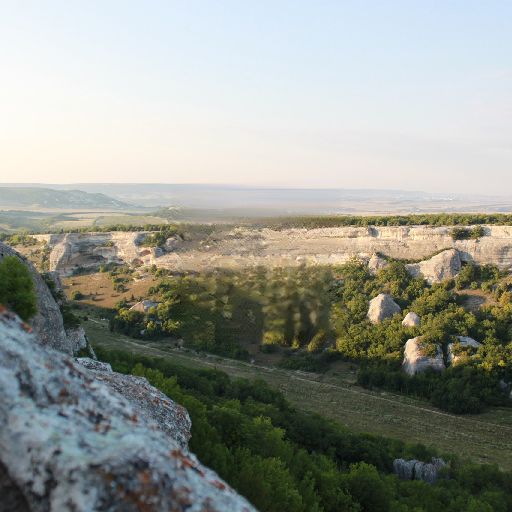} 
&   \includegraphics[valign=m]{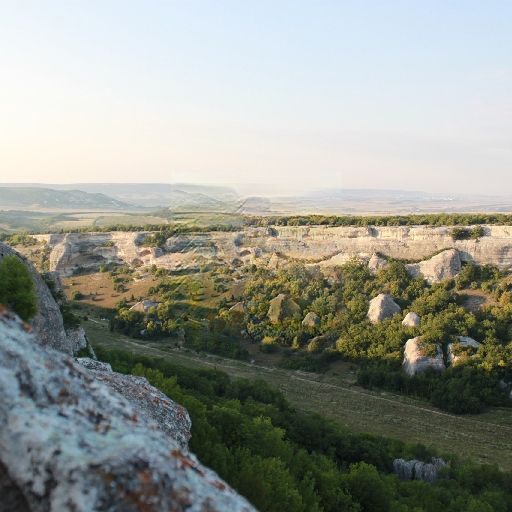}\\
\addlinespace[2pt]
\rothead{\centering \begin{math}{1024 \times 1024}\end{math}} 
&   \includegraphics[valign=m]{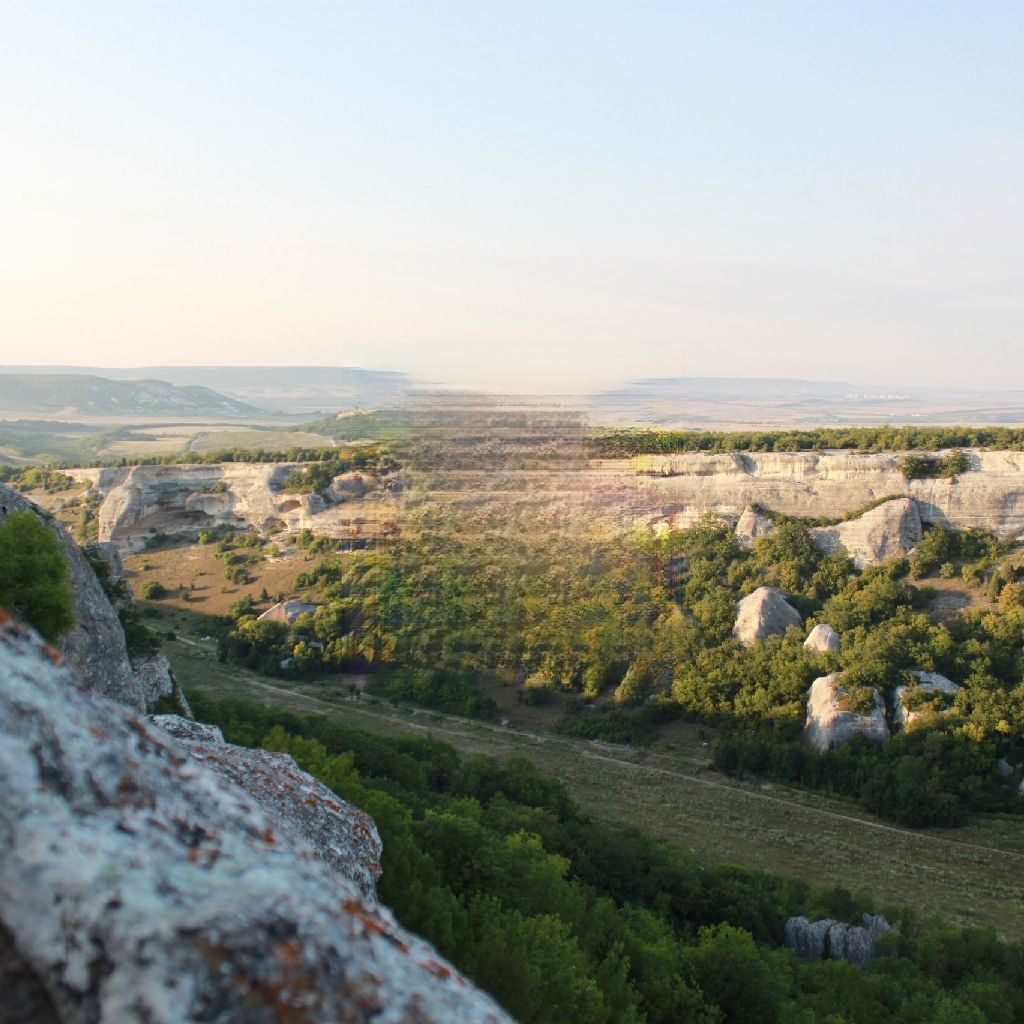}
&   \includegraphics[valign=m]{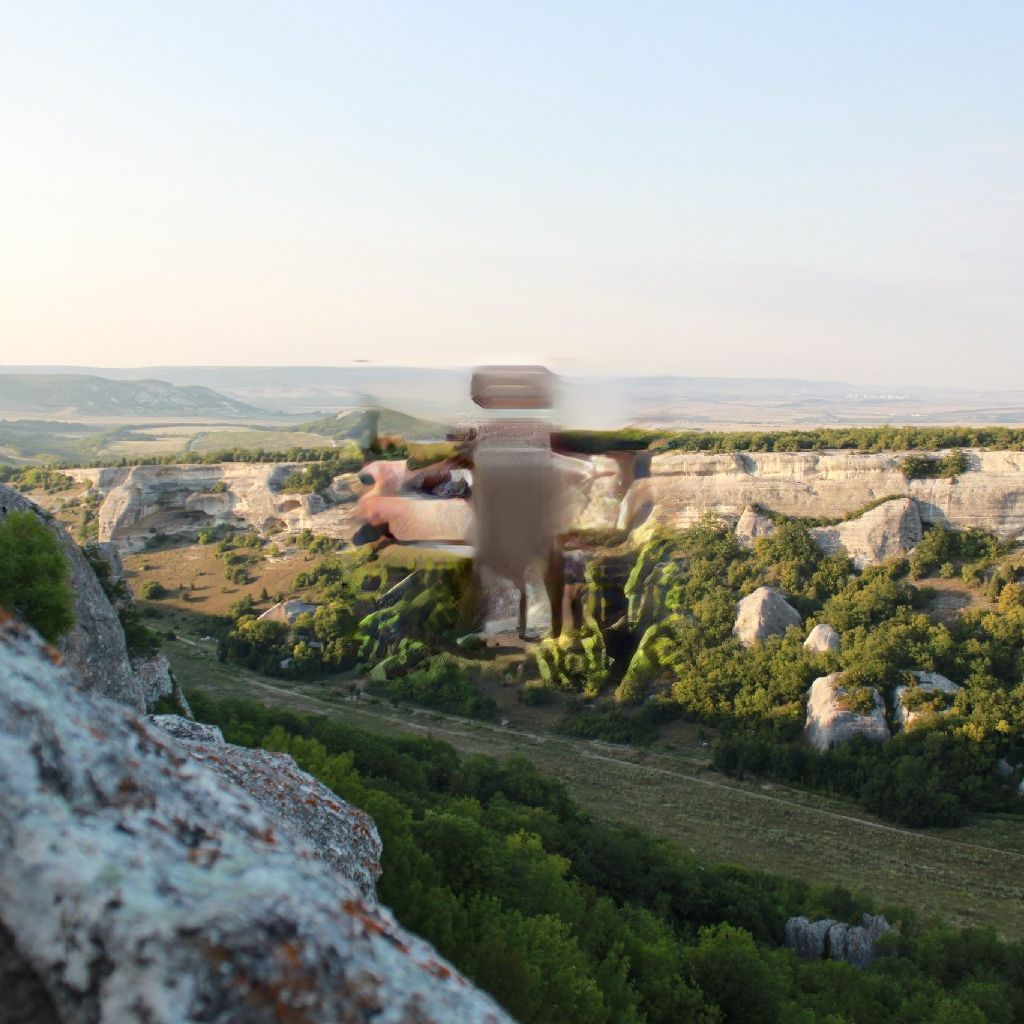}   
&   \includegraphics[valign=m]{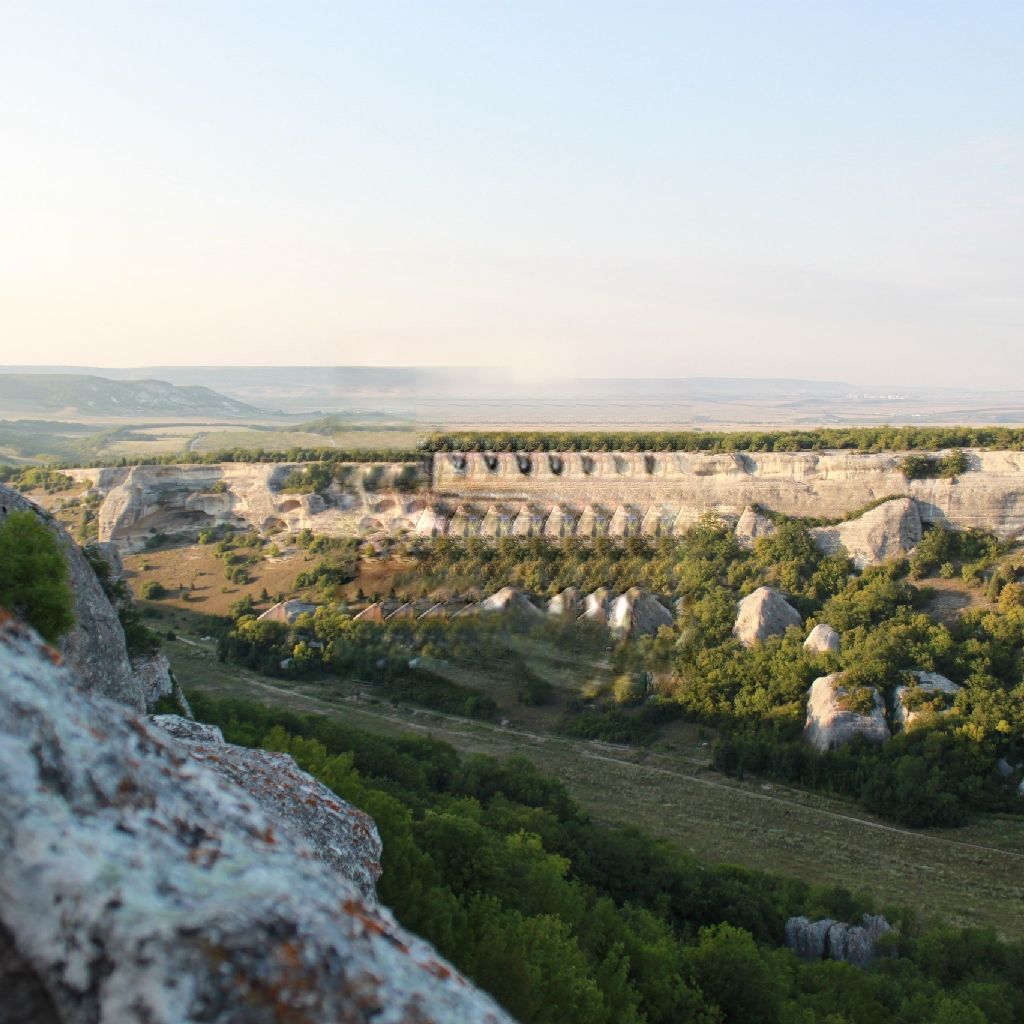}
&   \includegraphics[valign=m]{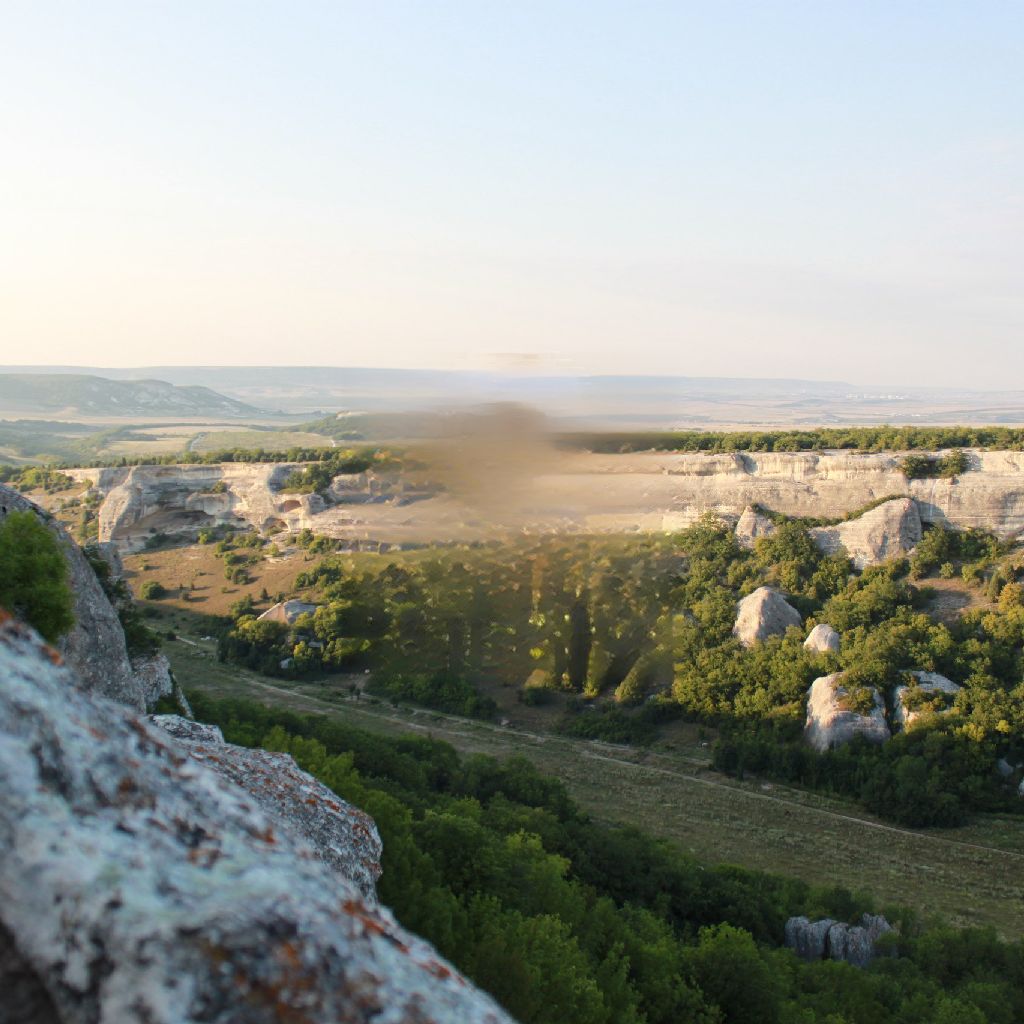}
&   \includegraphics[valign=m]{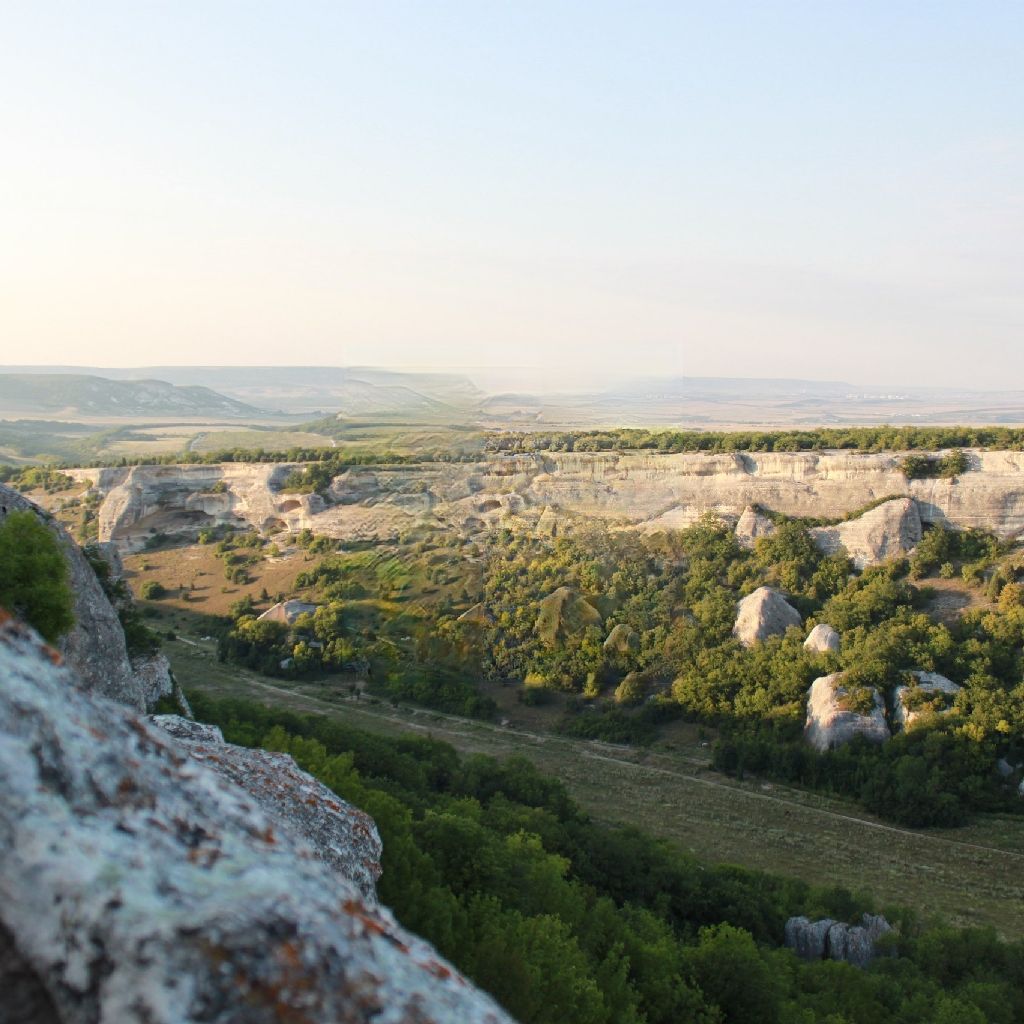}\\
\textbf{} &
\textbf{DFNet} &
\textbf{DeepFillv2} &
\textbf{HiFill} &
\textbf{ProFill} &
\textbf{Ours}
\end{tabularx}
\caption{Methods outputs at different resolutions}
\label{fig1}
\end{figure}

In recent years, the progress of neural networks has led to the development of deep inpainting methods. Neural-network methods are strongly tied to the resolution at which they are trained, owing to the lack of receptive field. Most models have an input size less than or equal to 512 pixels. As a result, they are unable to handle images of arbitrary shape — for instance, those in interactive image-processing tools. When the resolution increases, serious artifacts appear in the models. Fig.~\ref{fig1} shows several examples.

\noindent In this article, we describe a method that can restore images regardless of resolution. It uses the coarse-to-fine approach, restoring the image structure at low resolution and the texture at high resolution. Also, to improve texture filling, we propose using shifts of the original image that fill the hole and that artificially expand the receptive field by the shift amount. Our approach theoretically works for any inpainting method without retraining. 

\section{Related Work}

The solution to image inpainting problem can take the classical approach of choosing the most suitable patch \cite{FragmentBased,Criminisi,PatchMatch} from the image and sequentially filling the hole. Such methods are good at filling the texture component but poor at filling the structural component. In recent years, the development of neural networks has led to the creation of various deep inpainting methods \cite{Irregular,Genv1,DFNet}. Such algorithms avoid using external memory and operate only on the basis of knowledge gained during training. They are better than classical algorithms at restoring an image’s structural features. 

Adding to the difficulty of training neural-network methods is that the solution to the inpainting problem is nonunique. Thus, formulating the most suitable loss function for training is difficult. Deep-neural-network features are the best way to evaluate hole-filling quality \cite{Comparison}. 

Also, the appearance of generative adversarial networks (GANs) \cite{GANs} formed the basis for creating generative image-filling methods \cite{globally,Irregular,Genv1,Genv2,HiFill,ProFill}, which use adversarial loss as one component of their loss functions. 

\subsection{Gated Convolutions and Contextual Attention Module}
In \cite{Genv2}, researchers proposed replacing some of the neural network’s classical convolutions with gated convolutions, an extension of partial convolutions \cite{Irregular}. They also introduced a contextual-attention module, which is a neural-network analog of patch-based algorithms for filling image areas. Their method employs a GAN. Instead of a high-computational-cost contextual-attention module, we propose common shifts as a means of filling texture. 

\subsection{Deep Fusion Network}
The authors of \cite{DFNet} created a fusion block, which allows the network to, at its output, alpha blend each pixel in accordance with the predicted alpha map. They implemented the U-Net \cite{Ronneberger_2015} architecture in a non-generative-adversarial manner, using the high-level features of VGG16 \cite{VGG16} as loss functions. Our network implements a pretrained DFNet in the first stage, yielding the structural component in low resolution. We avoided a fusion block in our refinement network, since it changes even the unmasked area.

\subsection{High-Resolution Inpainting}
In \cite{HiFill}, researchers proposed modified gated convolutions that decrease the computational complexity, reducing the number of weights. They also suggested splitting the image at high and low frequencies by subtracting the blurry version from the image and using the modified contextual-attention module for aggregation. They trained their refinement network on the small but full images. The goal of our refinement network, on the other hand, is only to restore texture, so we train it on small patches of the original image; when testing, we use the entire image. 

\section{Proposed Approach}

\subsection{Data Preparation}
For training, we selected all images from DIV2K \cite{DIV2K} and some from the Internet. From each image, we cut out three random largest squares. In total, the training sample contained 7,218 images, with an additional 1,650 for validation. We applied to each image a random irregular mask from \cite{Irregular}. Our testing used natural images with a square mask at the center. 

\begin{figure}[ht]
\includegraphics[width=\textwidth]{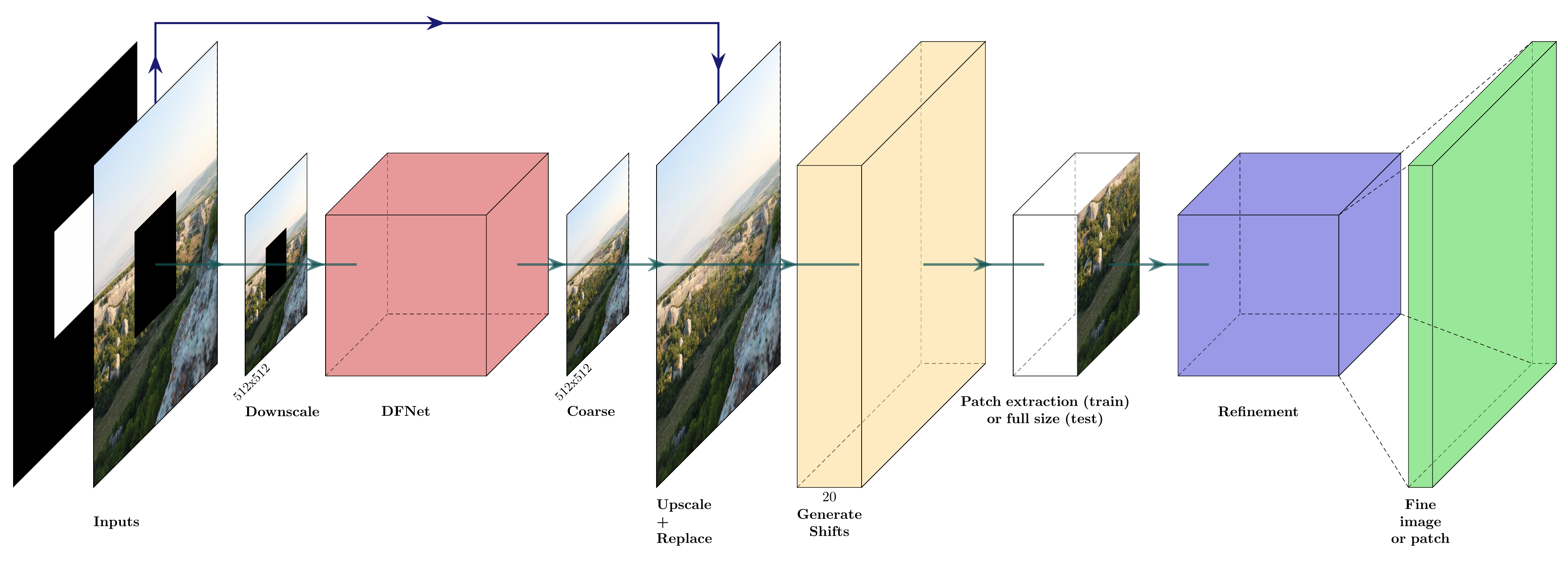}
\caption{Whole pipeline illustration} \label{fig2}
\vspace{0.45cm}
\includegraphics[width=\textwidth]{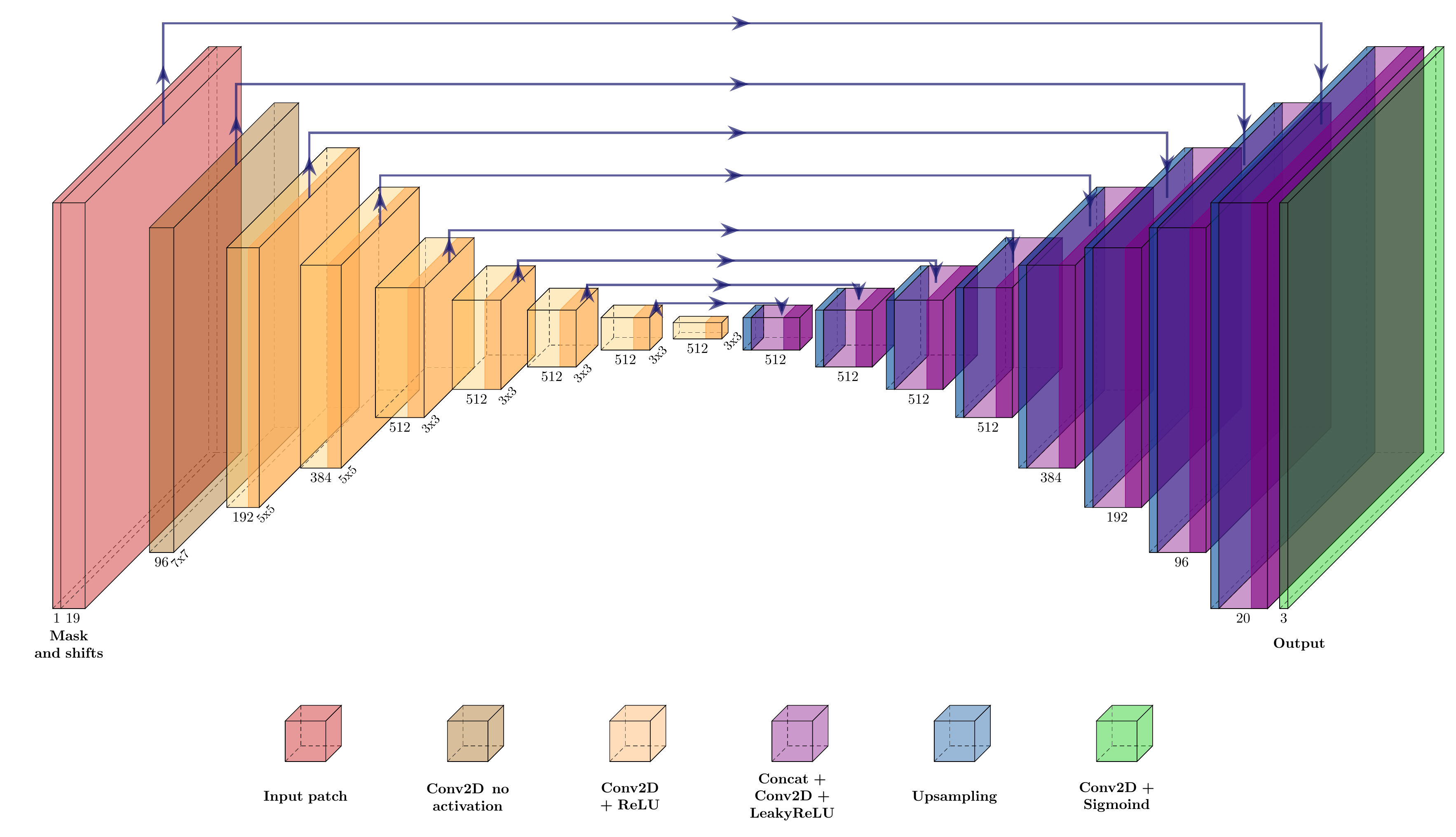}
\caption{Refinement network architecture} \label{fig3}

\end{figure}

\subsection{Whole Pipeline}
\subsubsection{Stage one}
The first stage of our algorithm restores the image structure at a low resolution. We initially downscale the image and mask to  \begin{math}{512\times512}\end{math}, then apply the pretrained DFNet  \cite{DFNet} to get the coarse result in low resolution. Next we perform an upscale and replace the known region with the corresponding region from the high-resolution image.

Finally we generate shifts of the original image in four directions: left, right, down, up. For our experiments, the shifts were 20\% of the image size. Note that during this process, we also recount the masks and mark the pixels in the open areas as invalid. Thus, the first stage yields a 20-channel image: five RGB images (main plus four shifts) and five masks. 

\subsubsection{Stage two}
The second stage restores the texture. To prevent the network from being attached to the structure and to increase training efficiency, we cut out random \begin{math}{512\times512}\end{math} patches from the input tensor of depth 20, with the condition that the masked area (from the main mask) is at least 10\% but not more than 90\% of the patch area. Note that during testing, we skip the patch extraction and simply transmit images in full resolution. The refinement network’s output is a fine filled result. Fig.~\ref{fig2} shows the pipeline. 

\subsection{Network Architecture}
The refinement-network architecture implements the U-Net  \cite{Ronneberger_2015} approach. (An illustration appears in Fig.~\ref{fig3}). Note that after each layer — except for the last one — we used Batch Normalization \cite{ioffe2015batch}. The figure shows encoder filter sizes; all decoder filters are \begin{math}{3\times3}\end{math}.

\subsection{Loss Function}
We trained our network in a non-generative-adversarial manner. Following \cite{DFNet} as a loss function, we used a linear combination:

\begin{equation}
\mathcal{L}=0.1 \cdot \mathcal{L}_{tv}+ 6.0 \cdot \mathcal{L}_{1}+ 0.1 \cdot \mathcal{L}_{p} + 240.0 \cdot \mathcal{L}_{s}
\end{equation}

\noindent Where for reference image \begin{math}\mathbf{I}\end{math} and predicted \begin{math}\hat{\mathbf{I}}\end{math}:

\noindent \begin{math}{L}_{tv}\end{math} — total variation distance in the masked area 

\noindent \begin{math}{L}_{1}\end{math} — distance which is calculated as

\begin{equation}
\mathcal{L}_{1} =\frac{1}{CHW}\left\|\mathbf{I}-\hat{\mathbf{I}}\right\|_{1}
\end{equation}

\noindent Where \begin{math}C,W,H\end{math} are the number of channels, width, and height respectively.

\noindent \begin{math}{L}_{p}, {L}_{s}\end{math} — Perceptual and Style Losses \cite{Johnson_2016}:

\begin{equation}
\mathcal{L}_{p}=\sum_{j \in J}\left\|\psi_{j}\left(\mathbf{I}\right)-\psi_{j}\left(\hat{\mathbf{I}}\right)\right\|_{1}
\end{equation}
\begin{equation}
\mathcal{L}_{s}=\sum_{j \in J}\left\|G_{j}\left(\mathbf{I}\right)-G_{j}\left(\hat{\mathbf{I}}\right)\right\|_{1}\end{equation}

\noindent Where \begin{math}J\end{math} is set of indices in VGG16, \begin{math}\psi_{j}\end{math} is \begin{math}j-th\end{math} feature layer. And \begin{math}G_{j}\end{math} is Gram matrix of \begin{math}j-th\end{math} feature layer.

\section{Experiments}
Our network training used the Adam \cite{kingma2014adam} optimizer with default settings. It took two days on two Nvidia Tesla P100 GPUs with a batch size of 18 images. Note that we trained only the network from the second stage. For optimization, we calculated the first stage’s output separately.

\begin{figure}[ht]
\centering
\setlength\tabcolsep{1pt}
\settowidth\rotheadsize{Radcliffe Cam}
\setkeys{Gin}{width=\hsize}
\begin{tabularx}{1.0\linewidth}{l XXXXXX}
&   \includegraphics[valign=m]{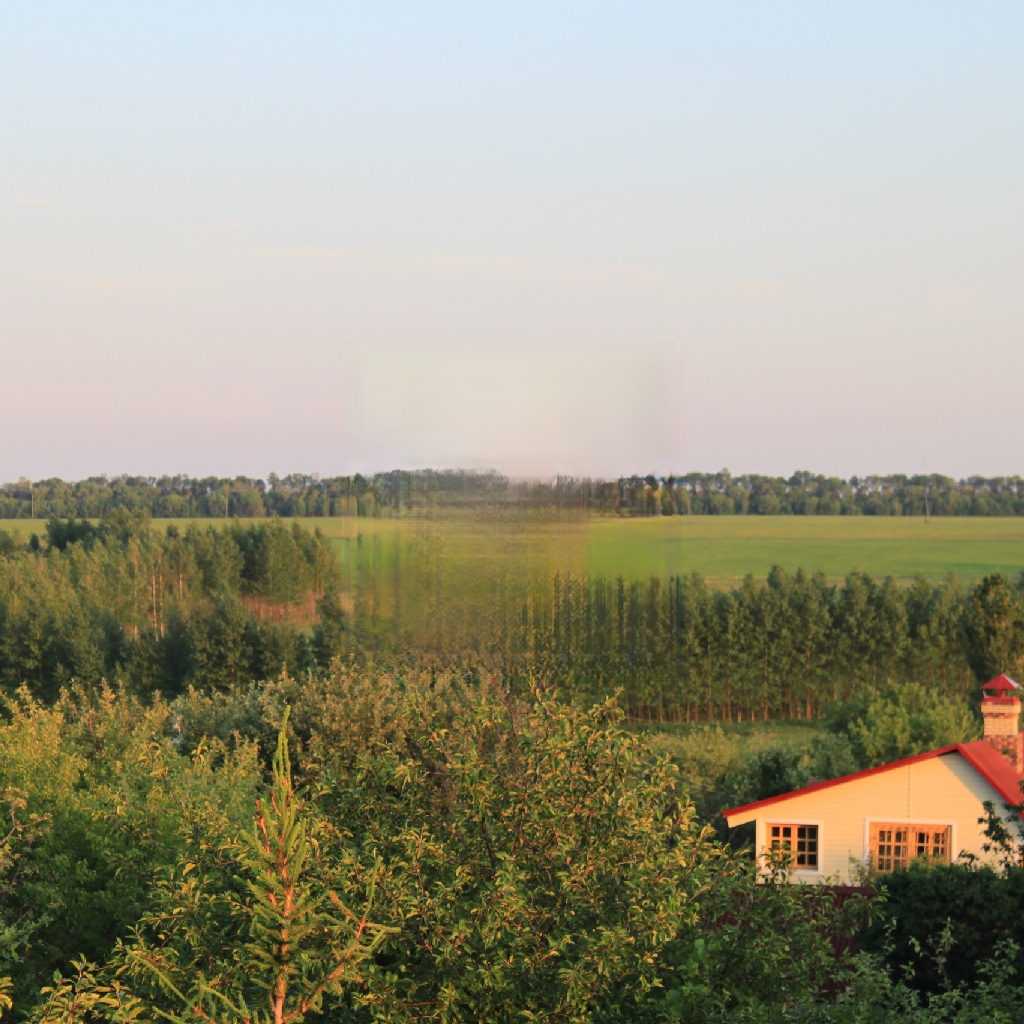}
&   \includegraphics[valign=m]{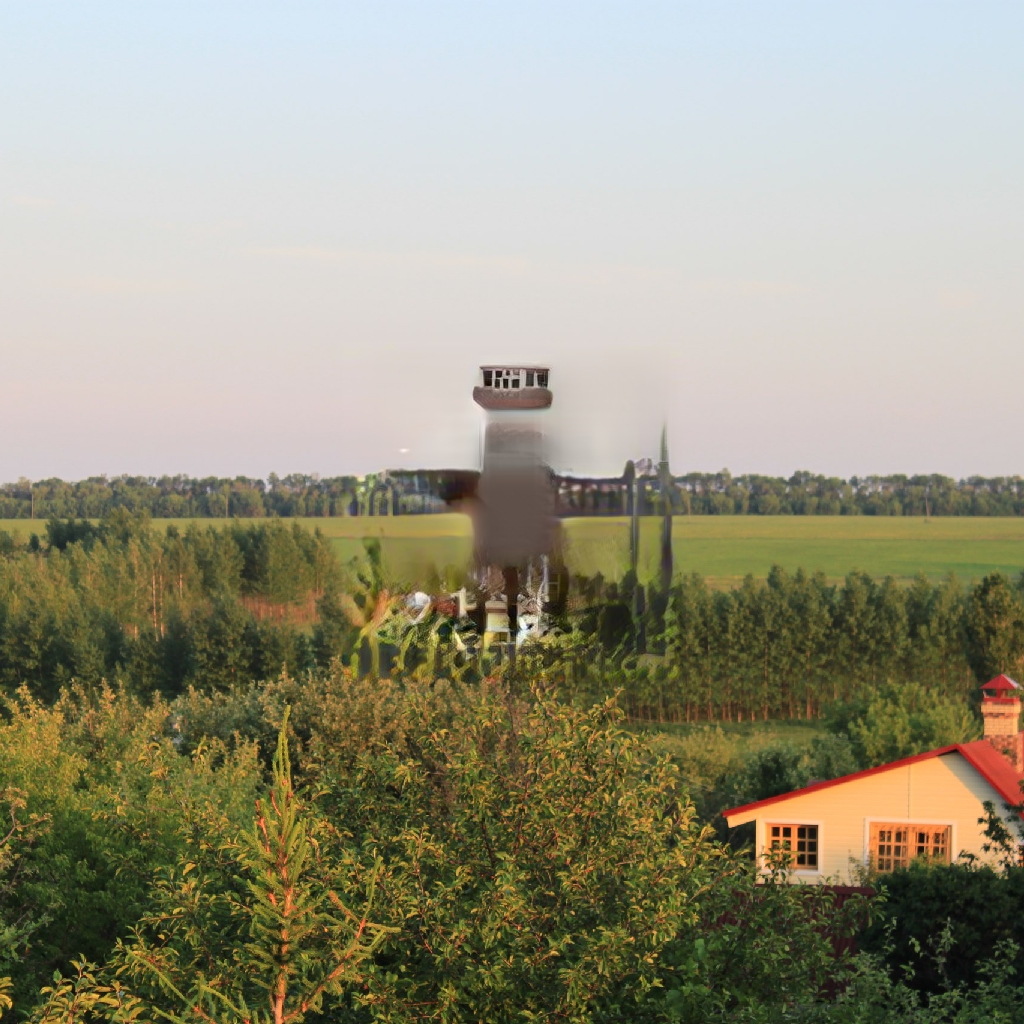}
&   \includegraphics[valign=m]{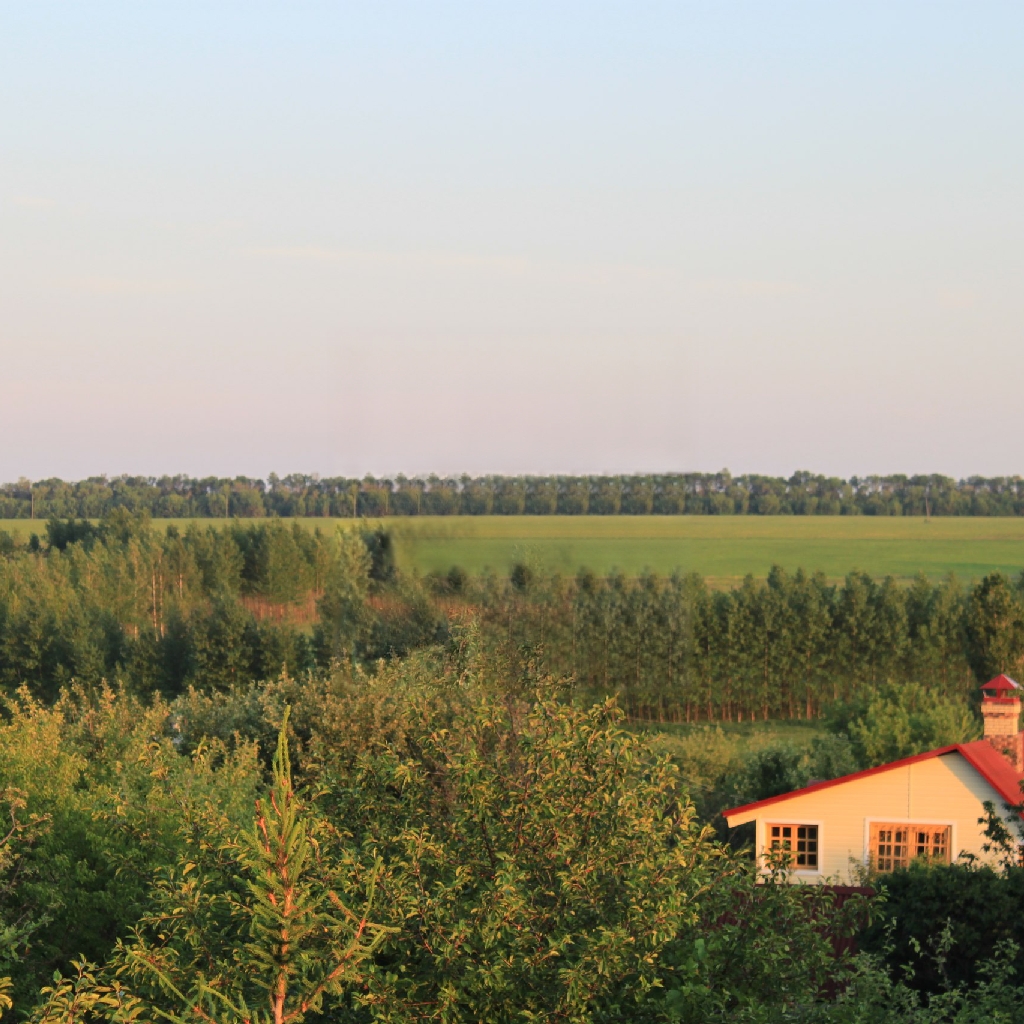}
&   \includegraphics[valign=m]{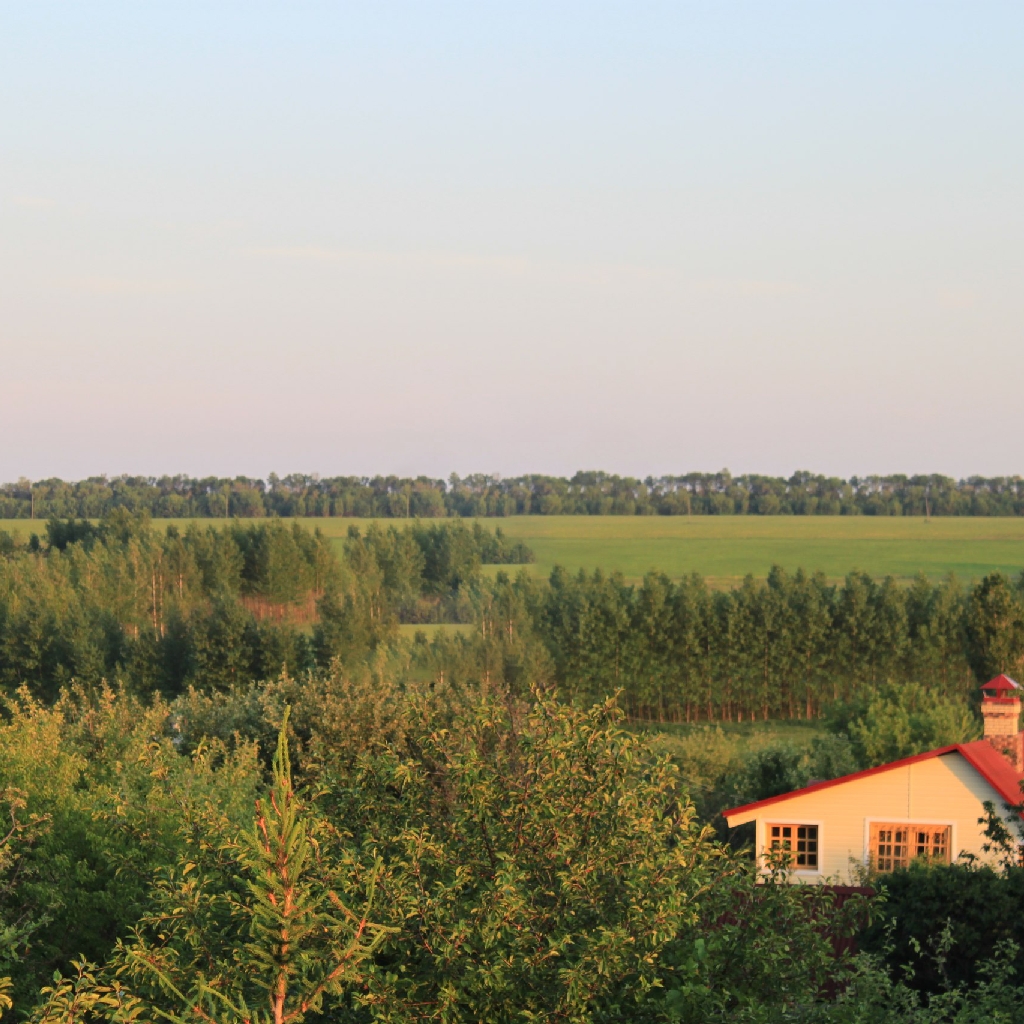}
&   \includegraphics[valign=m]{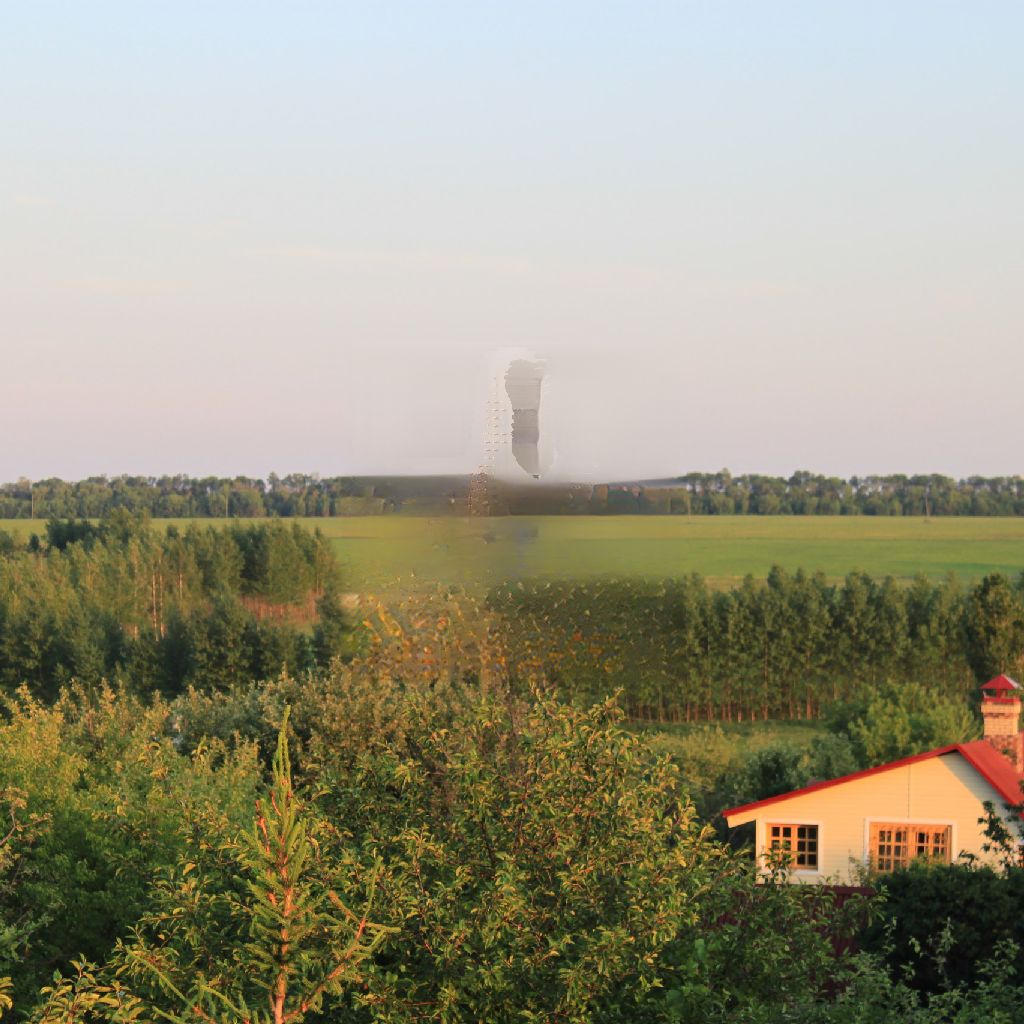}  
&   \includegraphics[valign=m]{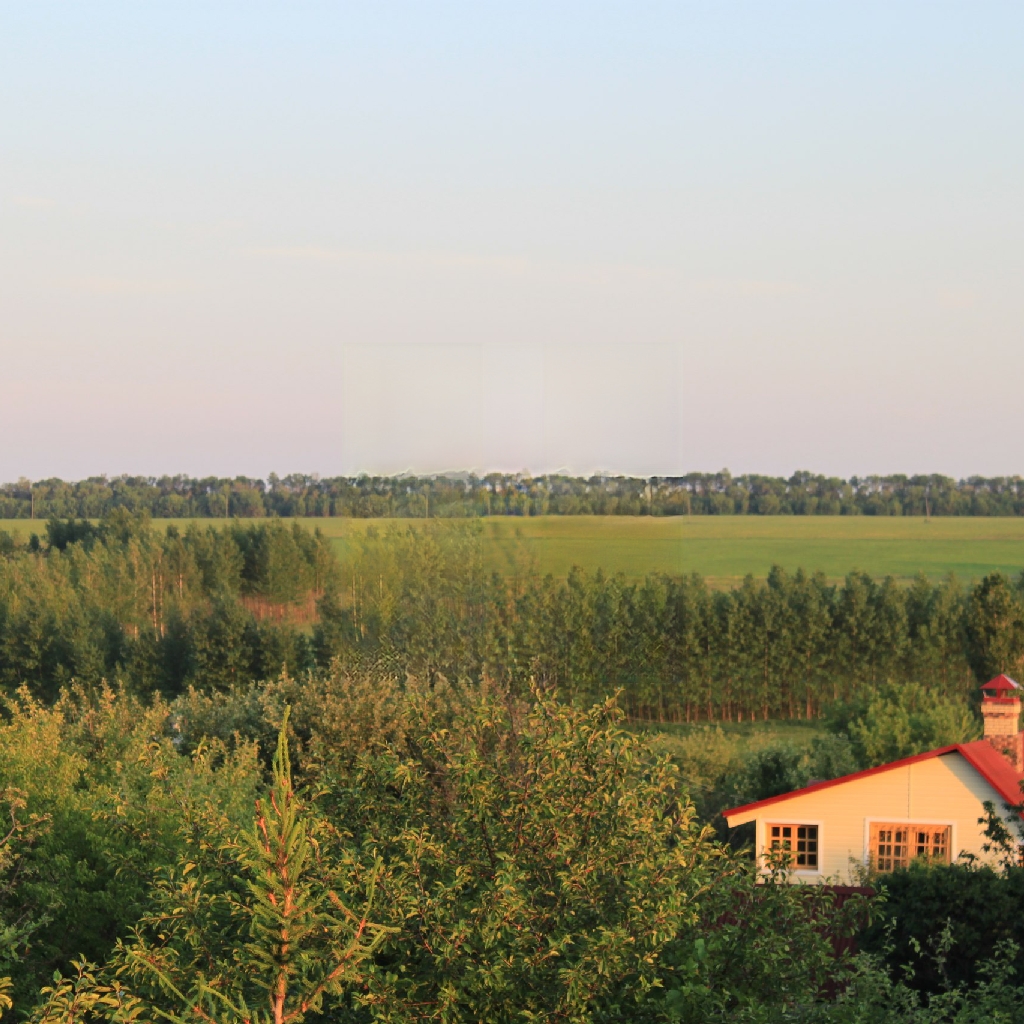}  \\
\addlinespace[2pt]
&   \includegraphics[valign=m]{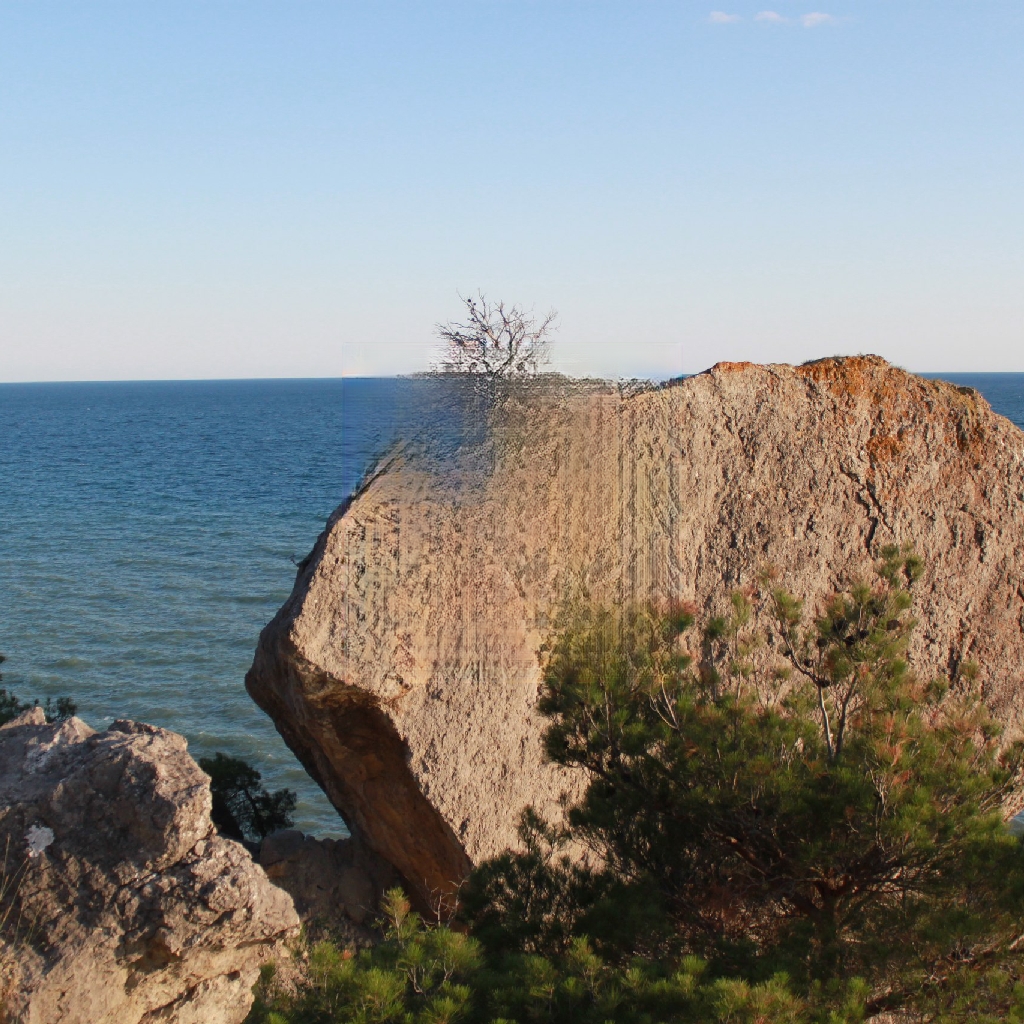}
&   \includegraphics[valign=m]{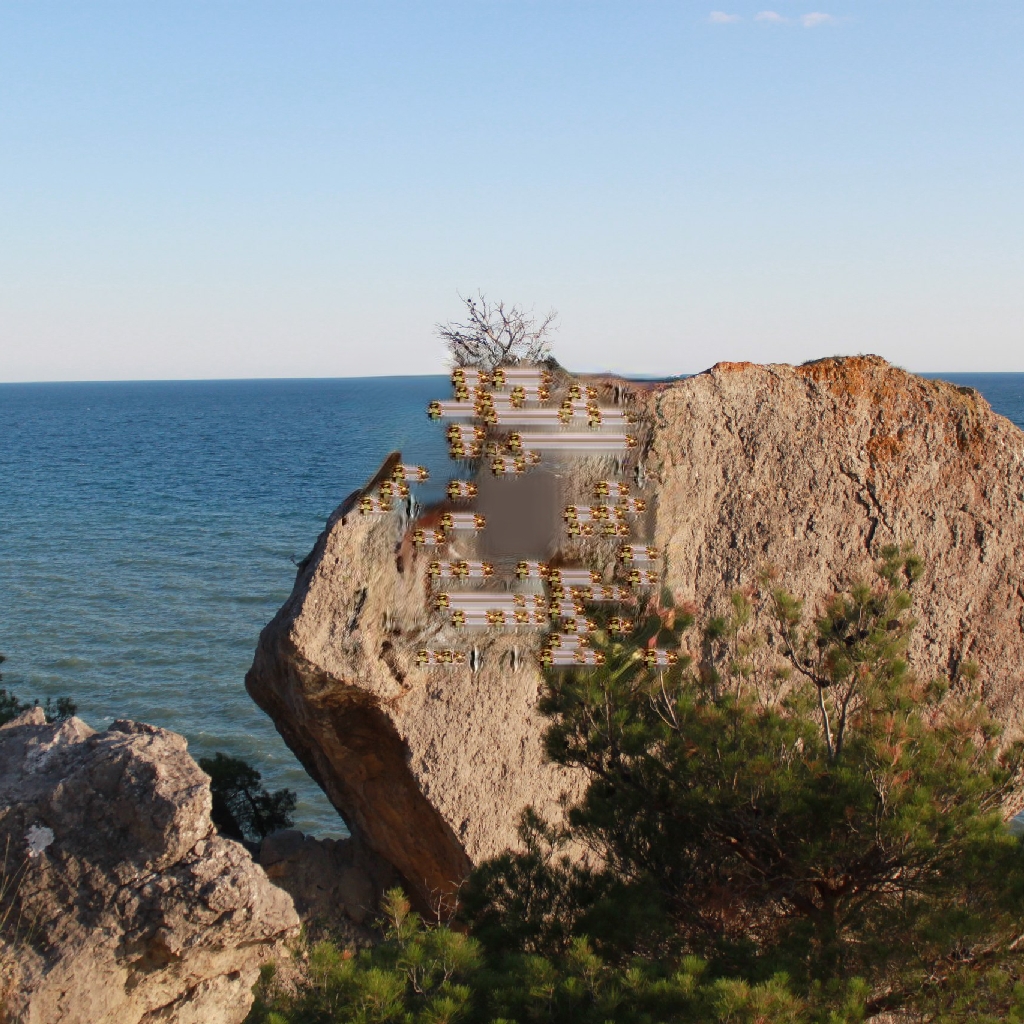}
&   \includegraphics[valign=m]{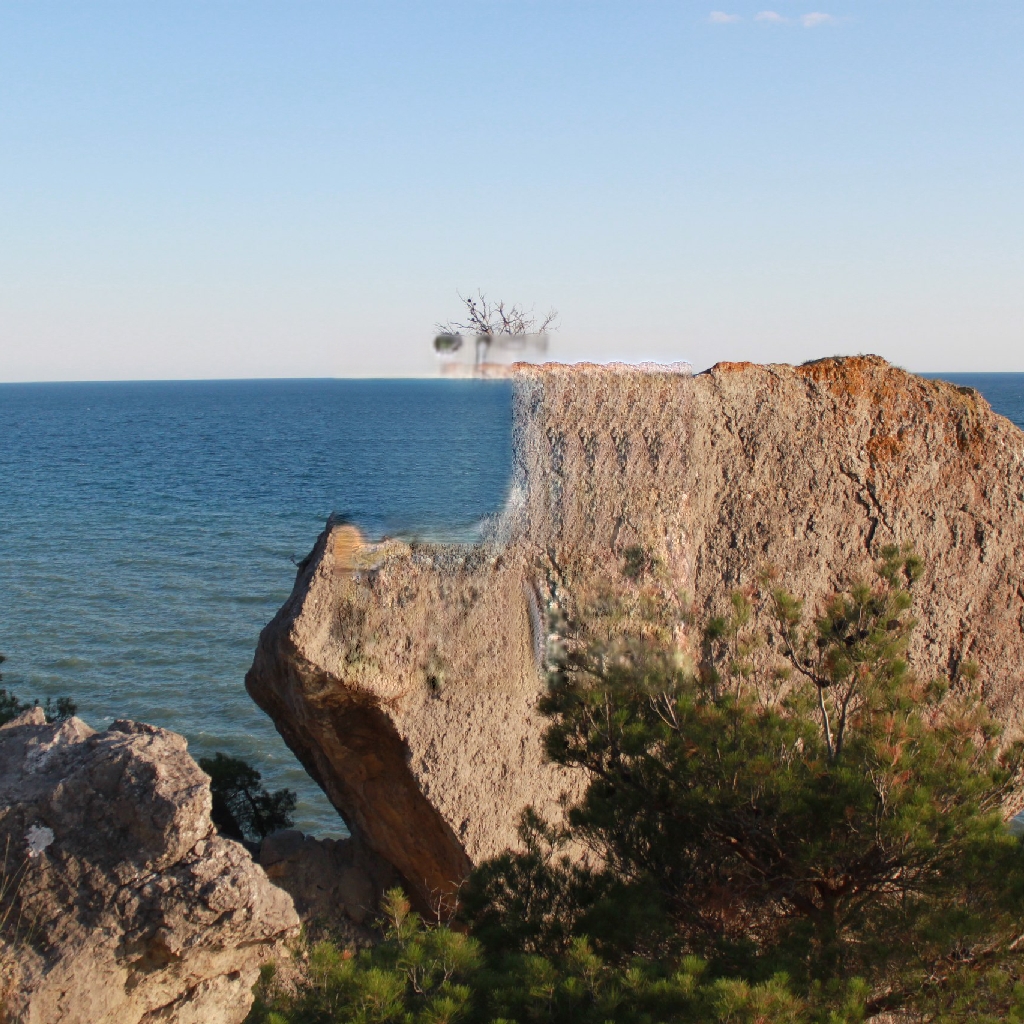}
&   \includegraphics[valign=m]{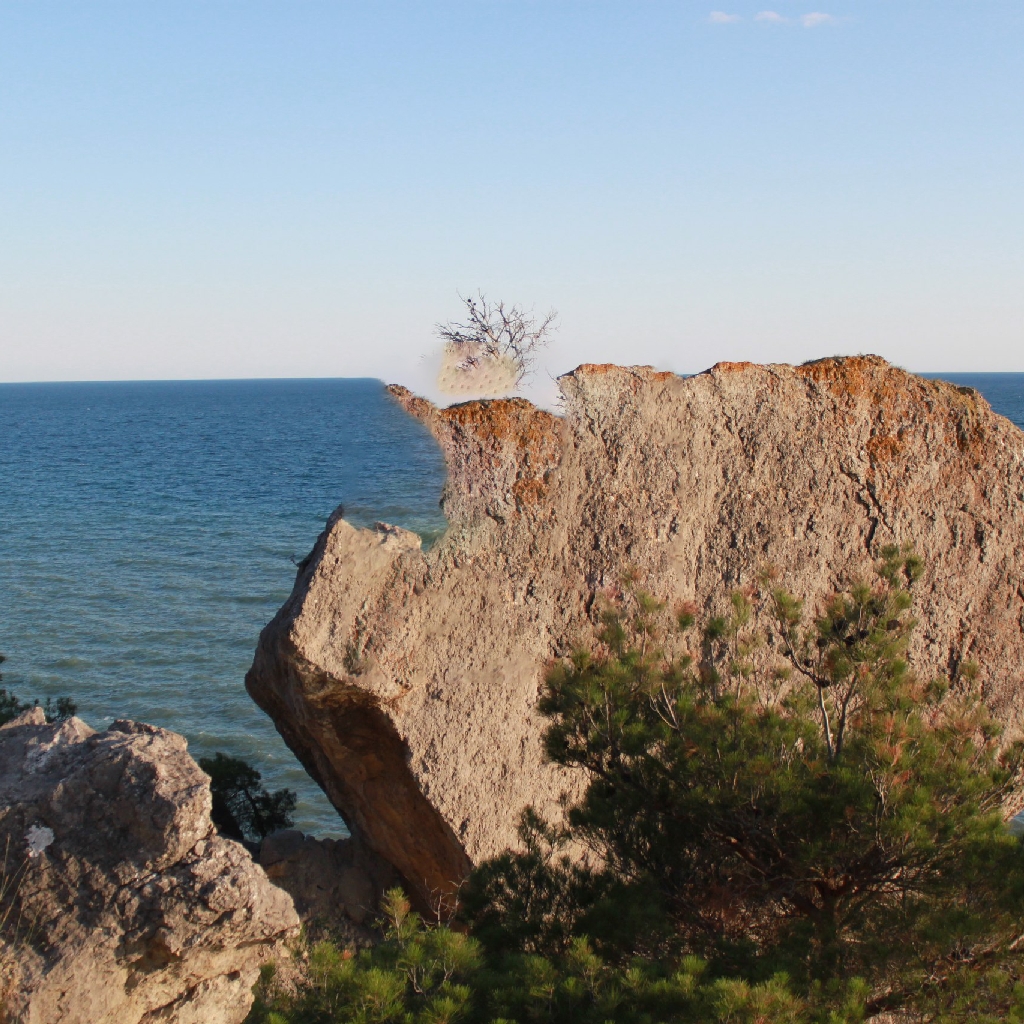}
&   \includegraphics[valign=m]{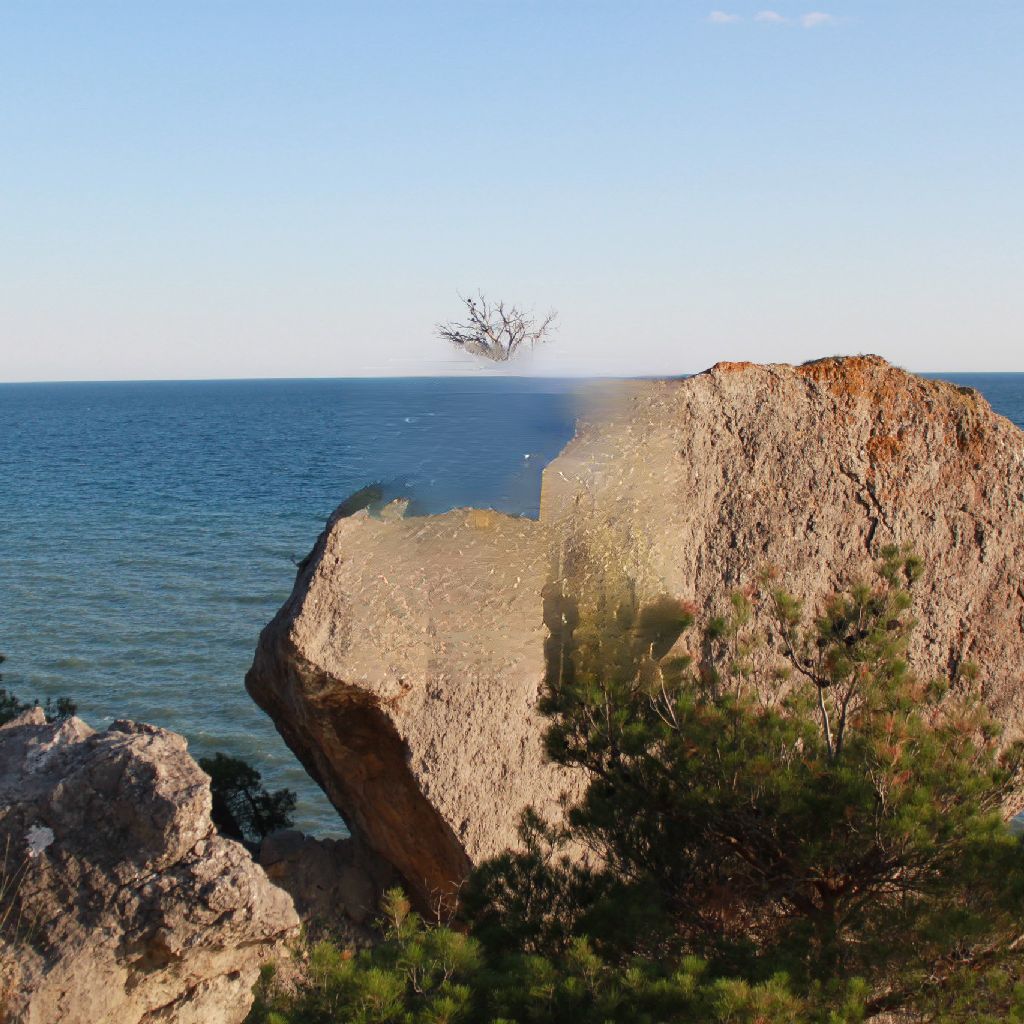} 
&   \includegraphics[valign=m]{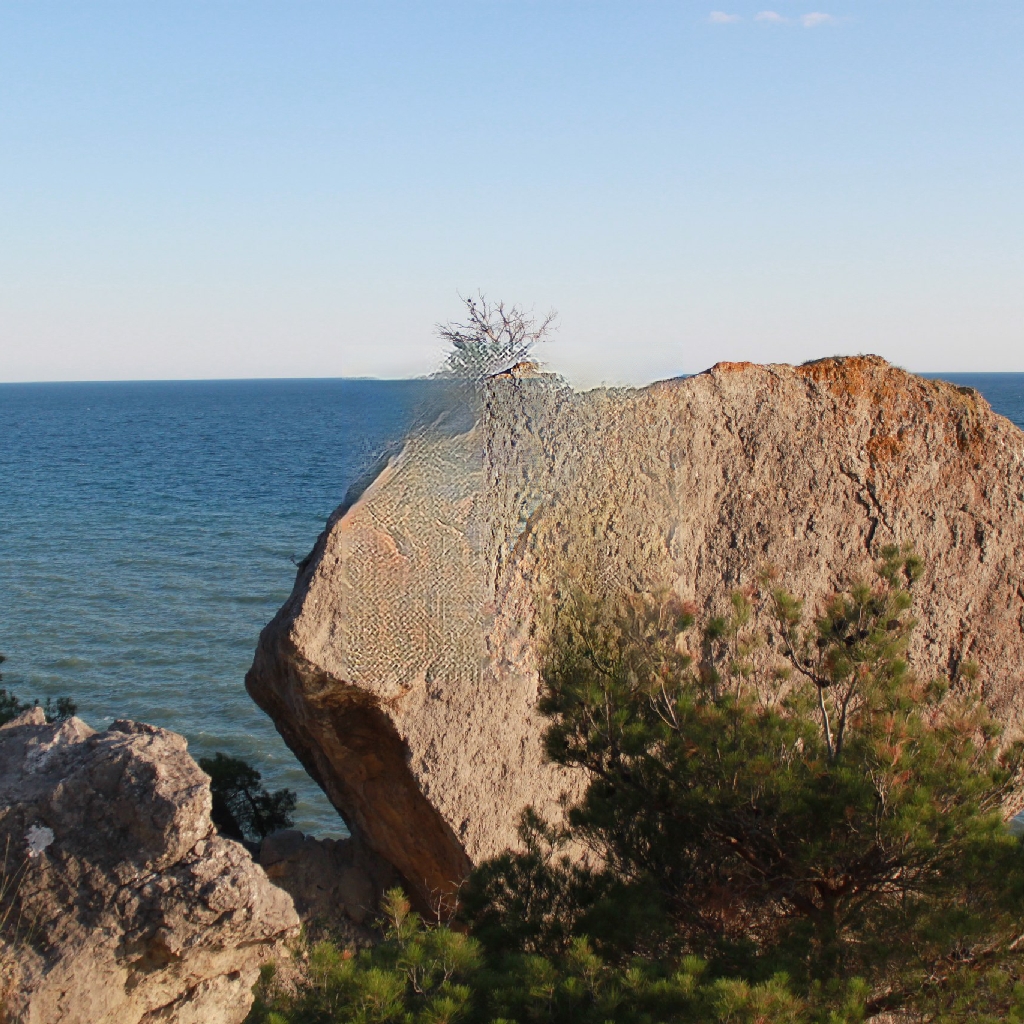}  \\
\addlinespace[2pt]
&   \includegraphics[valign=m]{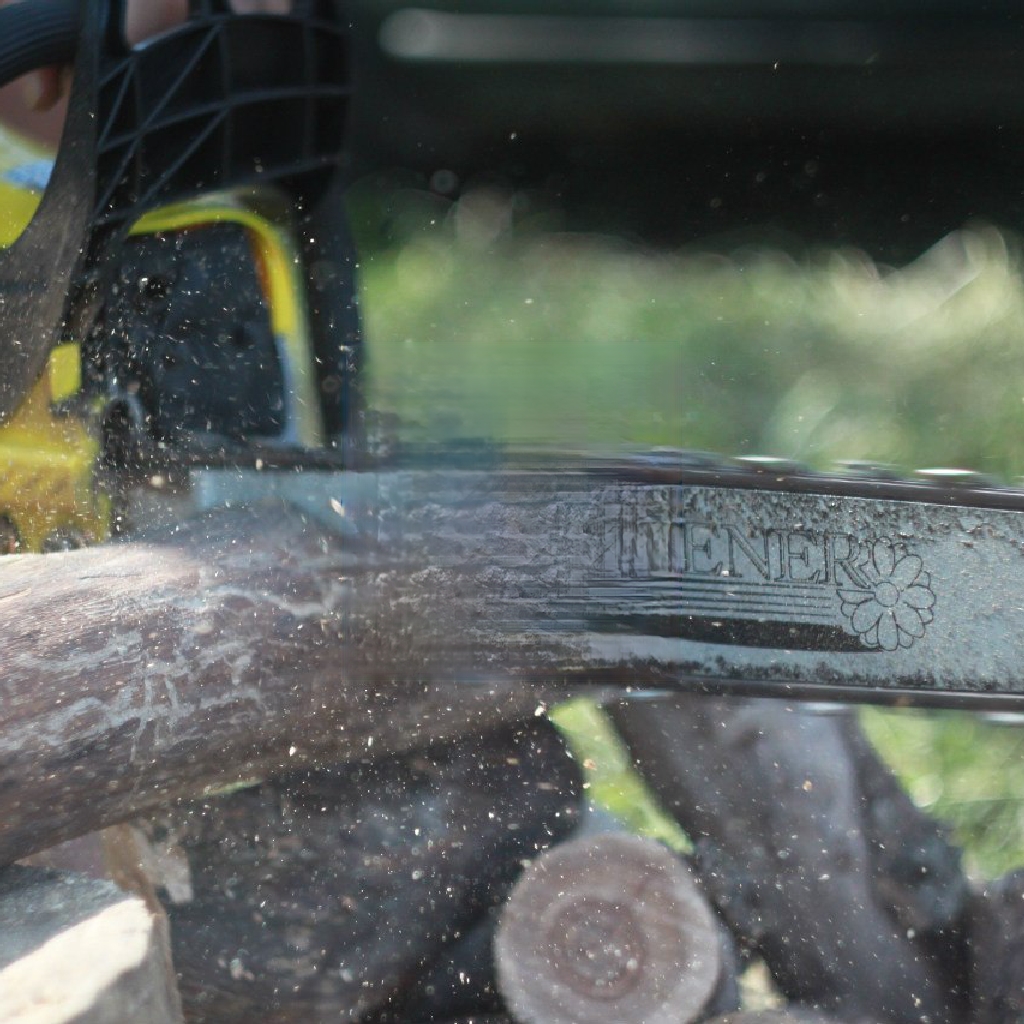}
&   \includegraphics[valign=m]{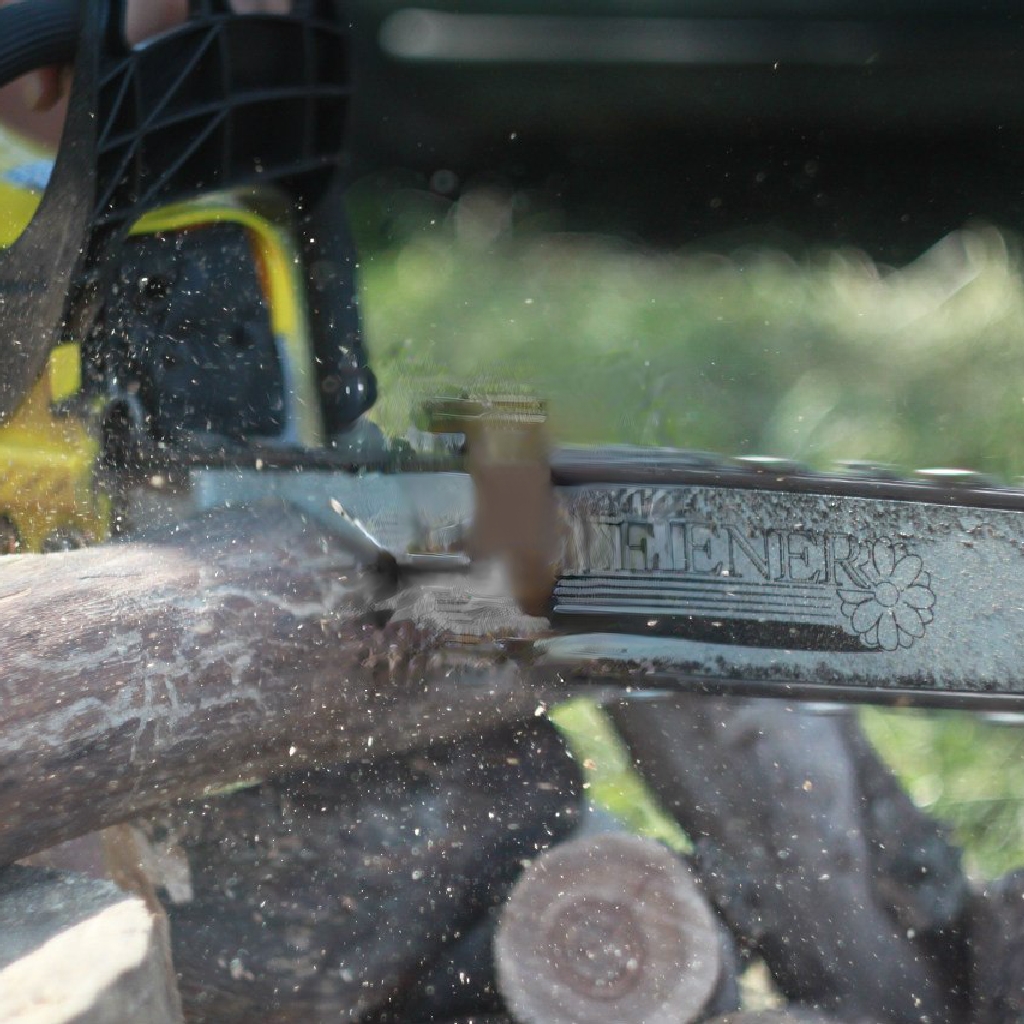}    
&   \includegraphics[valign=m]{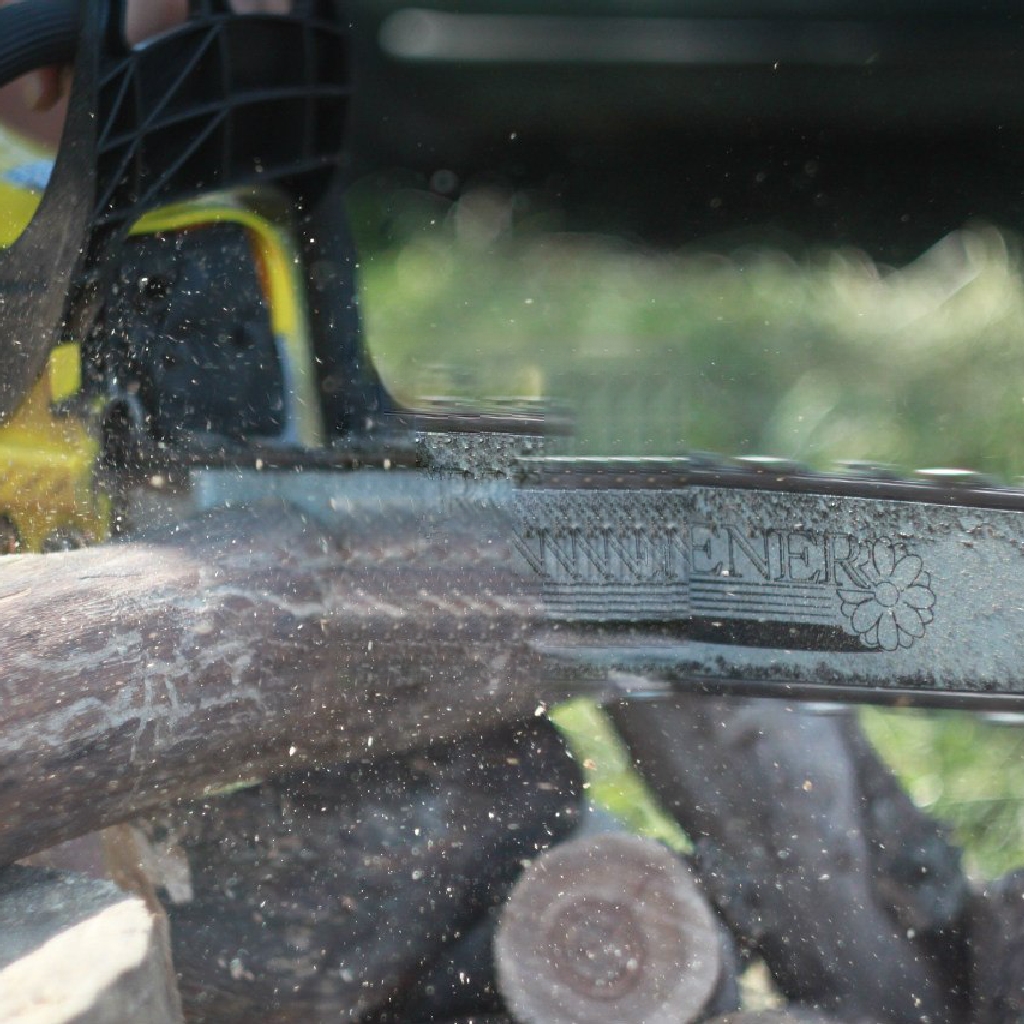}
&   \includegraphics[valign=m]{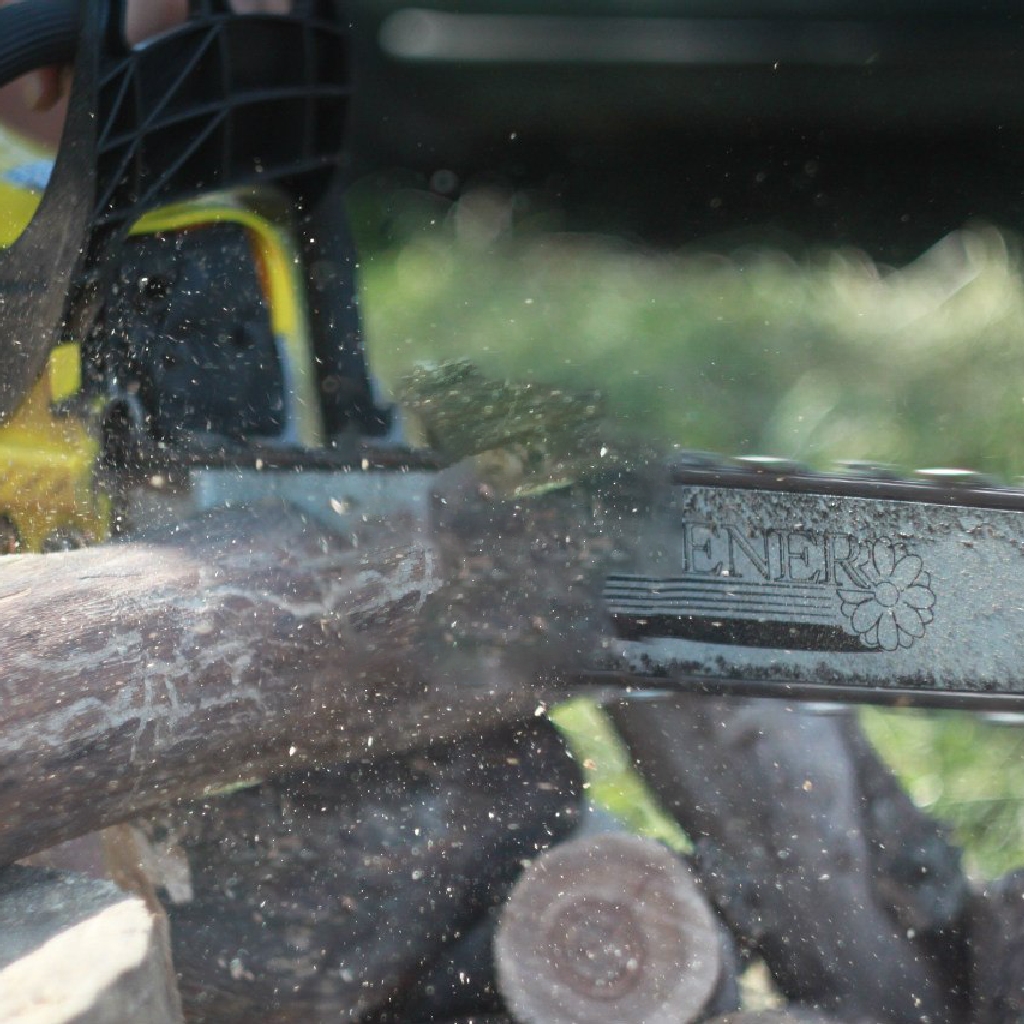}
&   \includegraphics[valign=m]{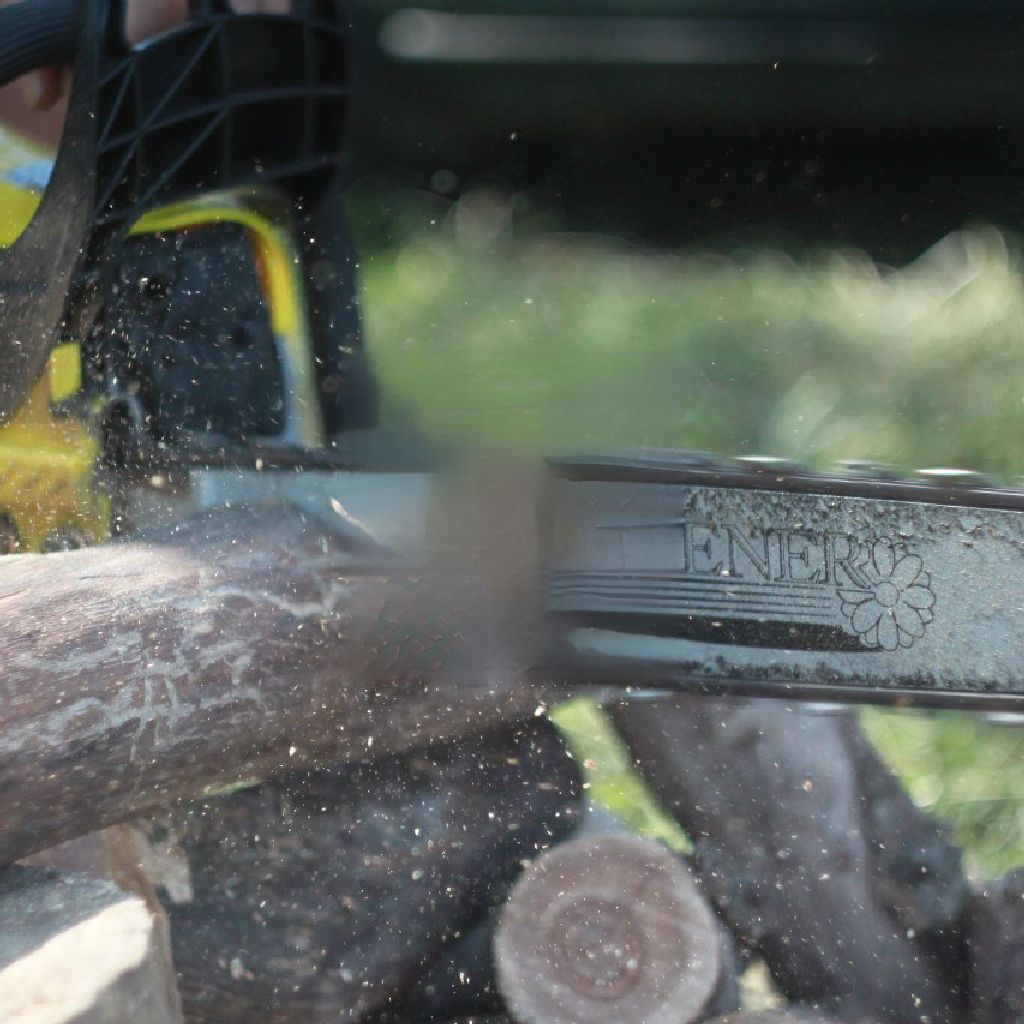}
&   \includegraphics[valign=m]{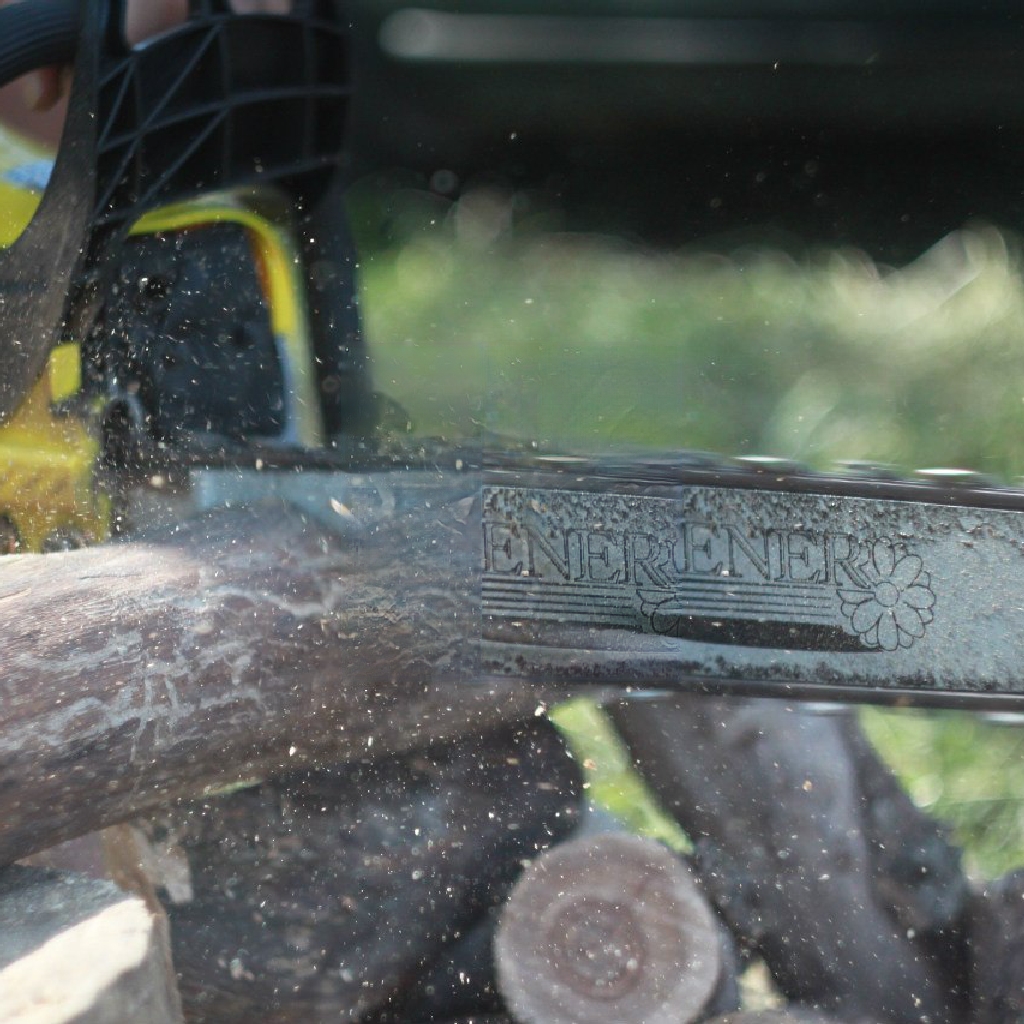}\\
\addlinespace[2pt]
&   \includegraphics[valign=m]{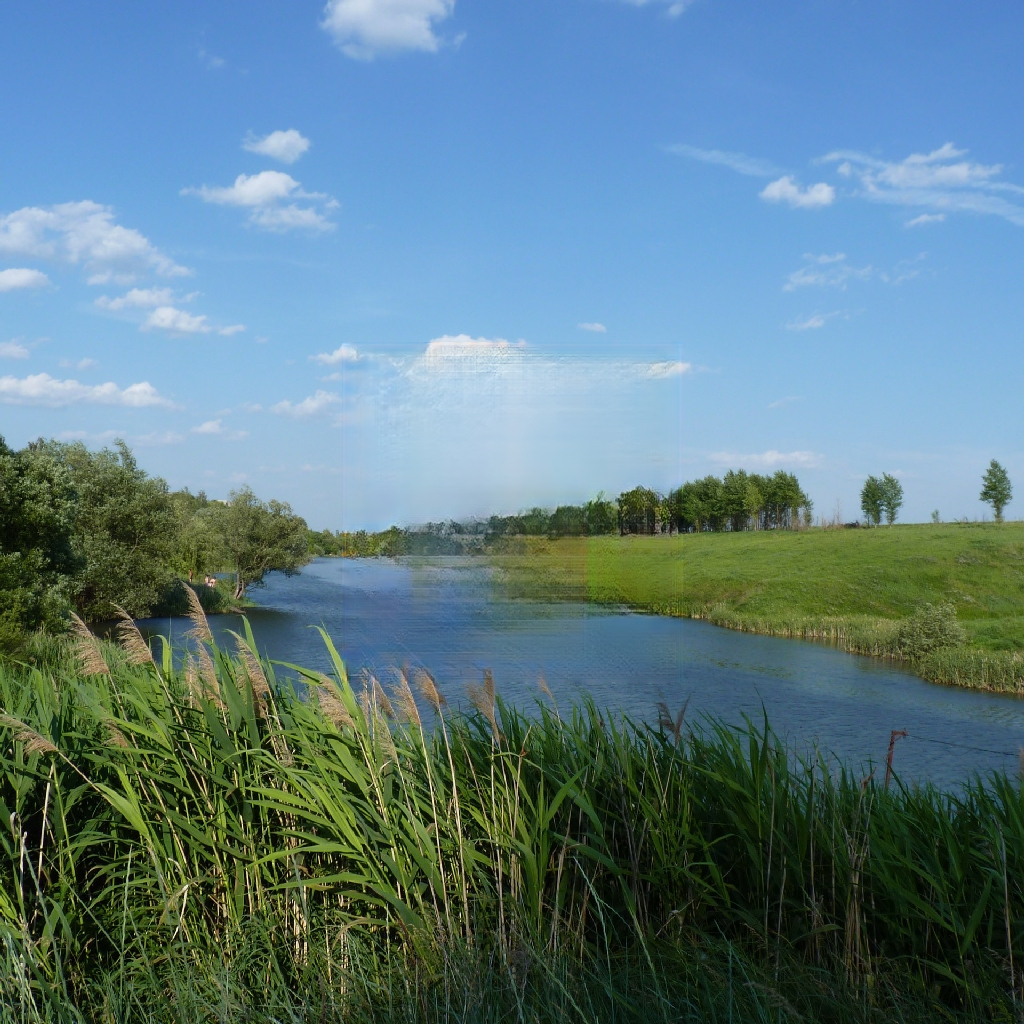}
&   \includegraphics[valign=m]{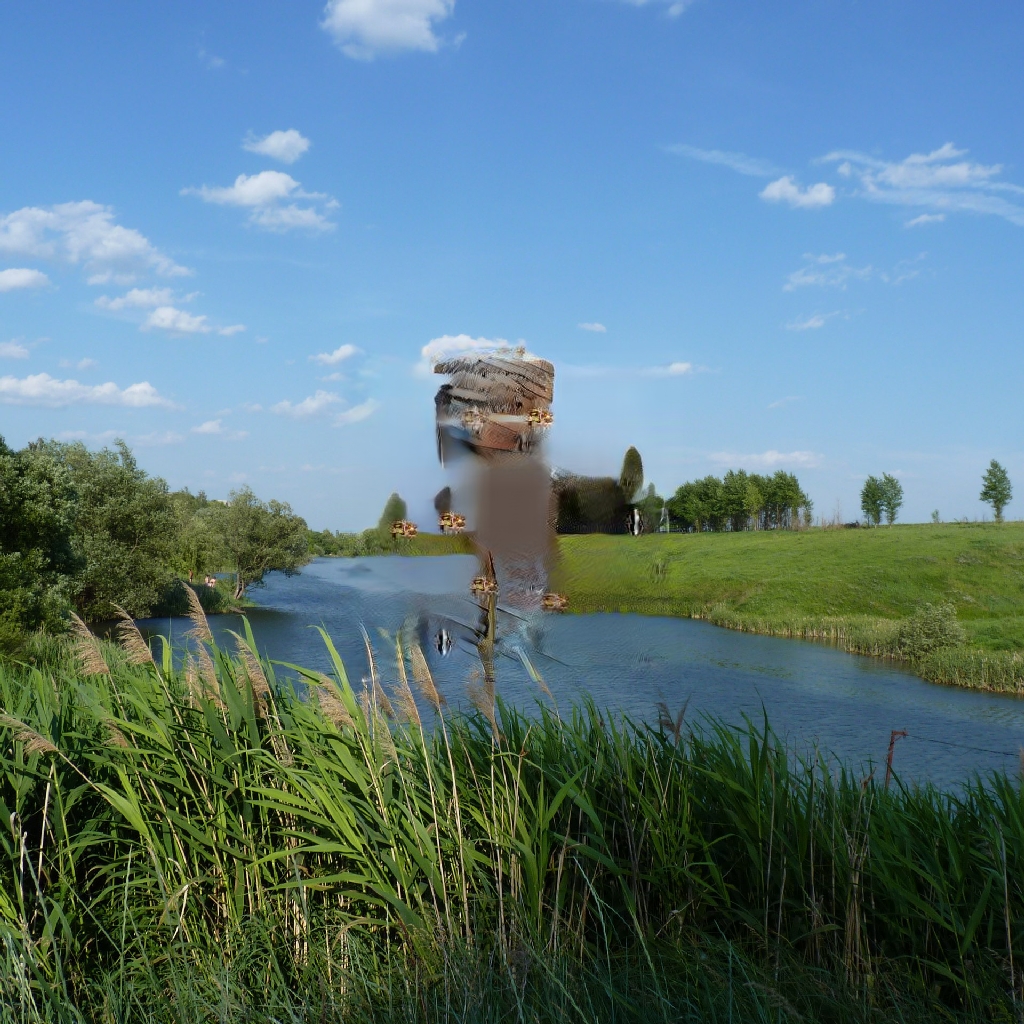}    
&   \includegraphics[valign=m]{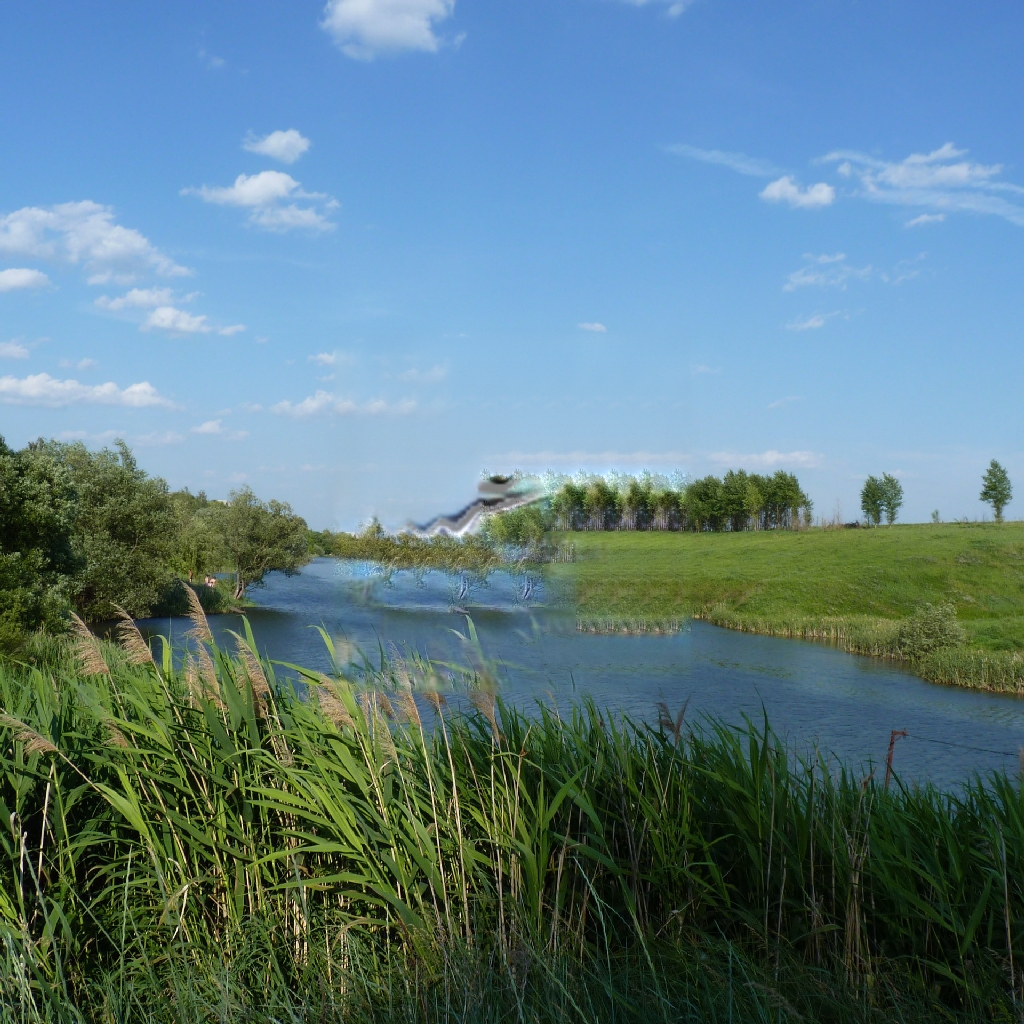}
&   \includegraphics[valign=m]{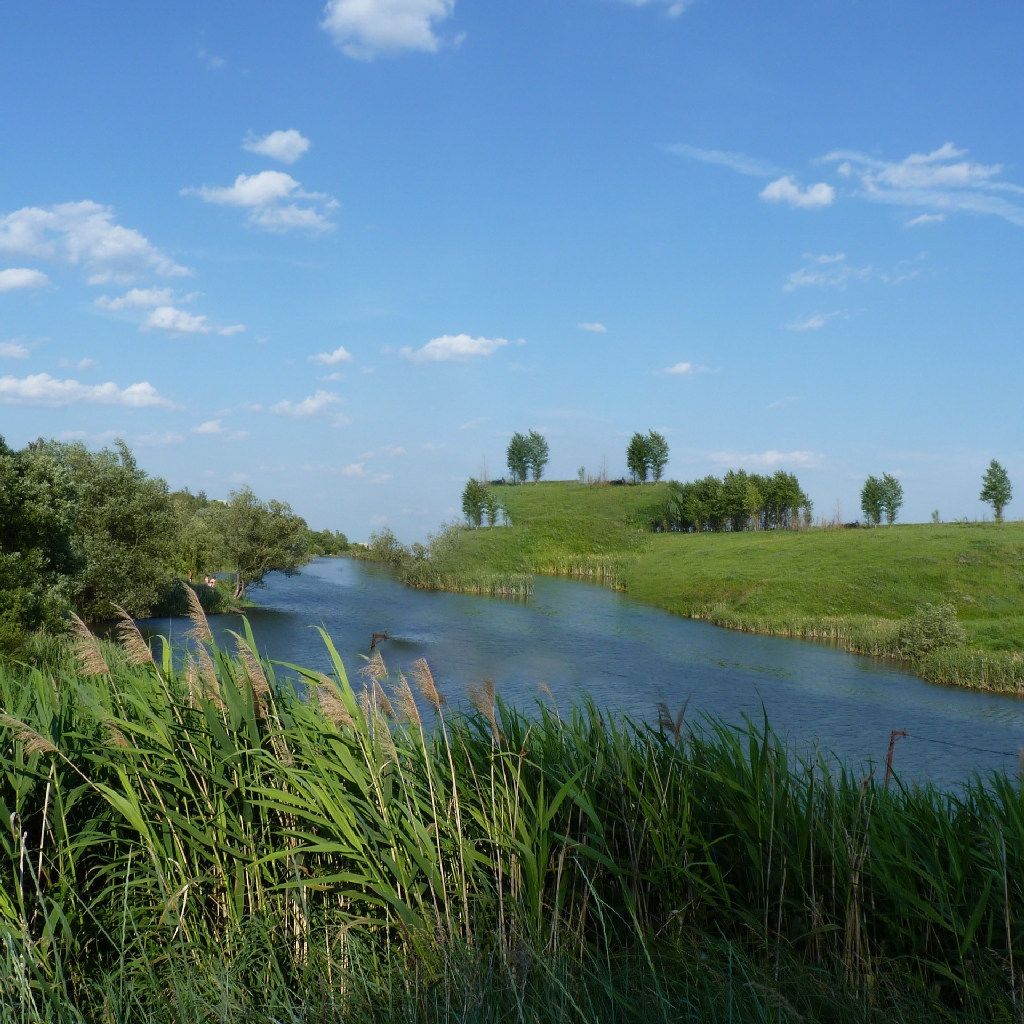}
&   \includegraphics[valign=m]{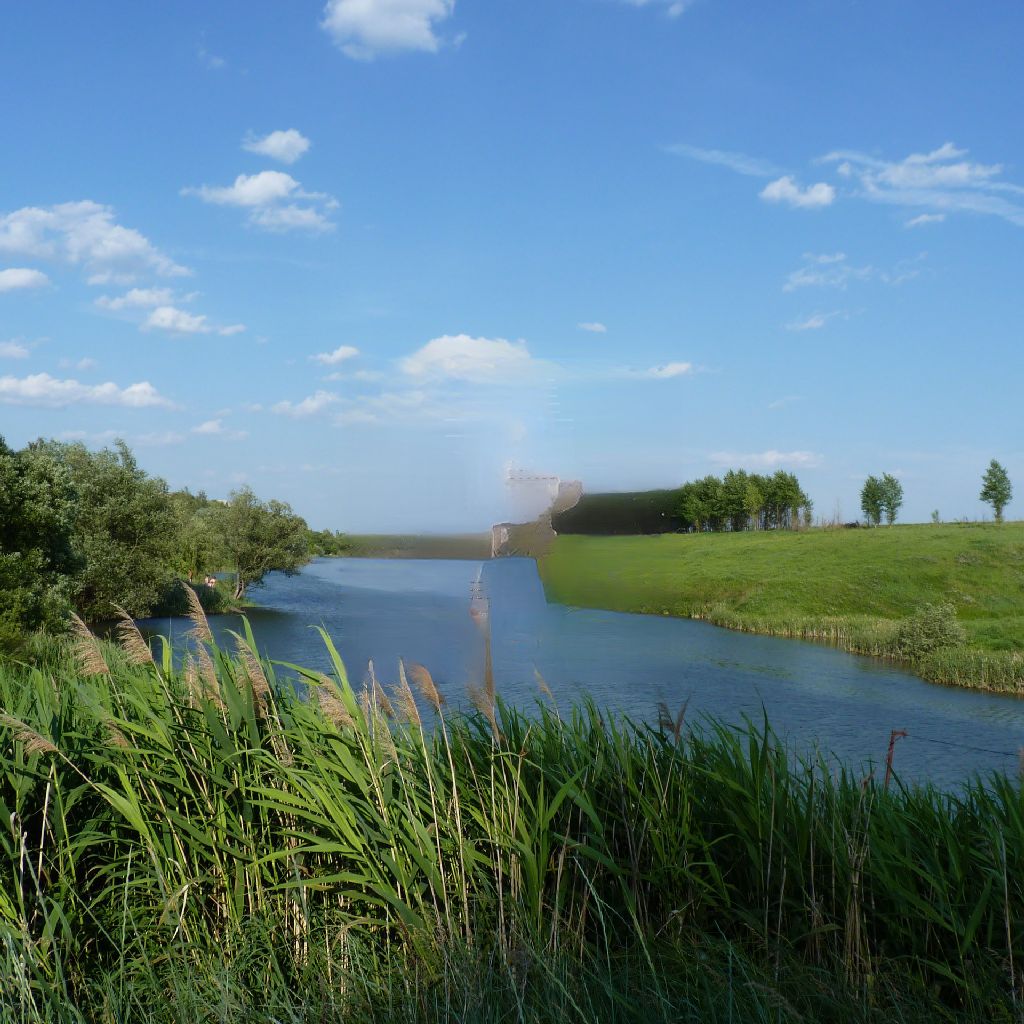}
&   \includegraphics[valign=m]{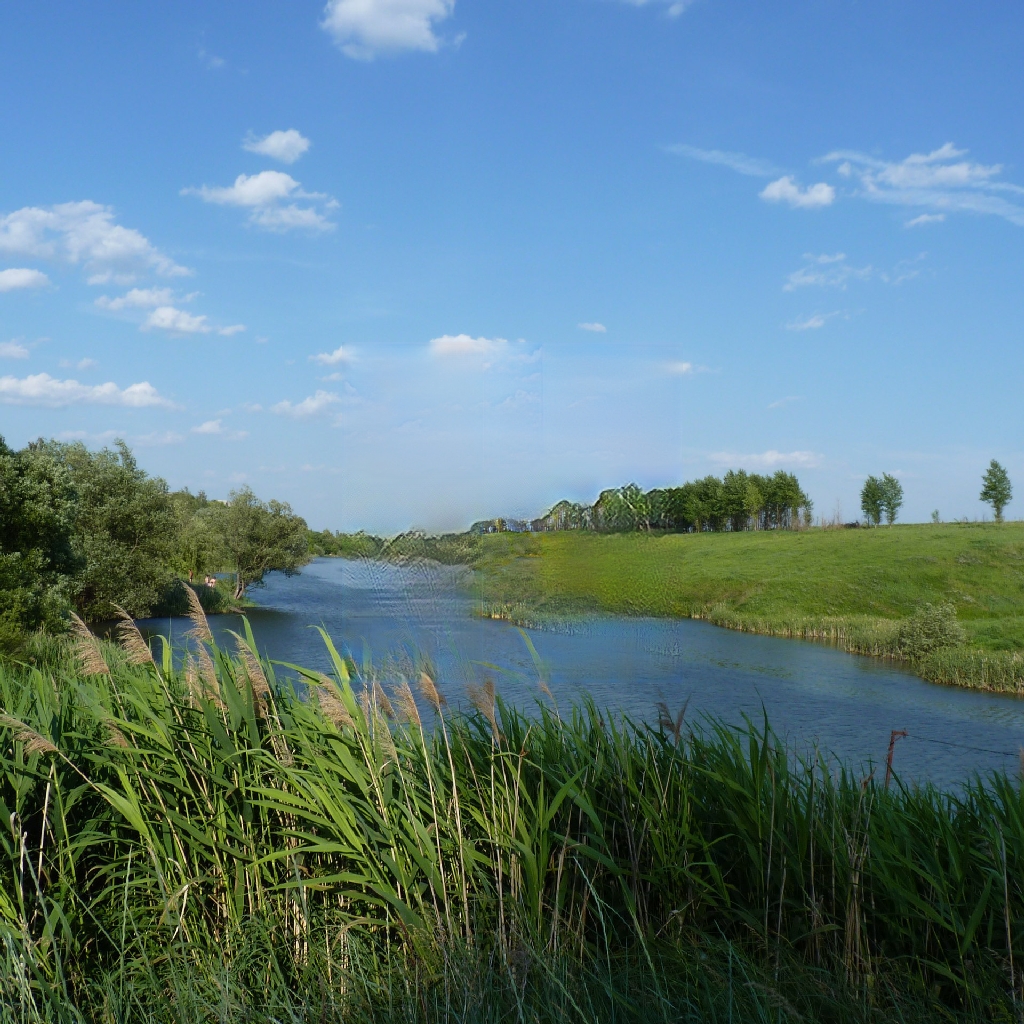}\\
\textbf{} &
\textbf{DFNet} &
\textbf{DeepFillv2} &
\textbf{HiFill} &
\textbf{Photoshop} &
\textbf{ProFill} &
\textbf{Ours}
\end{tabularx}
\caption{Methods outputs with \begin{math}{1024\times1024}\end{math} images}
\label{fig5}
\end{figure}

\subsection{Comparisons}
Fig.~\ref{fig1}, Fig.~\ref{fig5}, and Fig.~\ref{fig6} show examples of our work. Although the method functions almost identically at any resolution, we limited ourselves to \begin{math}{1024\times1024}\end{math}. More pictures and resolutions appear in the repository.

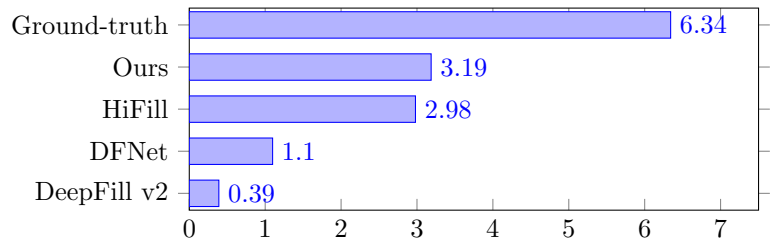
\begin{figure}
\centering
\begin{tikzpicture}[trim axis right]
\pgfplotsset{%
    width=0.75\textwidth,
    height=0.35\textwidth
}
\begin{axis}[
xbar, xmin=0, xmax=7.5,
symbolic y coords={%
    {DeepFill v2},
    {DFNet},
    {HiFill},
    {Ours},
    Ground-truth},
ytick=data,
nodes near coords,
nodes near coords align={horizontal},
ytick=data,
]
\addplot coordinates {
    (1.0979,{DFNet})
    (0.3896,{DeepFill v2})
    (2.9800,{HiFill})
    (3.1865,{Ours})
    (6.3418,Ground-truth)};
\end{axis}
\end{tikzpicture}
\caption{Subjective comparison scores (resolution 2048x2048)} \label{fig41}
\end{figure}

\begin{figure}
\centering
\begin{tikzpicture}[trim axis right]
\pgfplotsset{%
    width=0.75\textwidth,
    height=0.35\textwidth
}
\begin{axis}[
xbar, xmin=0, xmax=7.5,
symbolic y coords={%
    {DeepFill v2},
    {ProFill},
    {HiFill},
    {DFNet},
    {Ours},
    Ground-truth},
ytick=data,
nodes near coords,
nodes near coords align={horizontal},
ytick=data,
]

\addplot coordinates {
    (2.238,{DFNet})
    (0.437,{DeepFill v2})
    (1.613,{ProFill})
    (2.201,{HiFill})
    (3.212,{Ours})
    (6.402,Ground-truth)};
\end{axis}
\end{tikzpicture}
\caption{Subjective comparison scores (resolution 1024x1024)}
\label{fig42}
\end{figure}
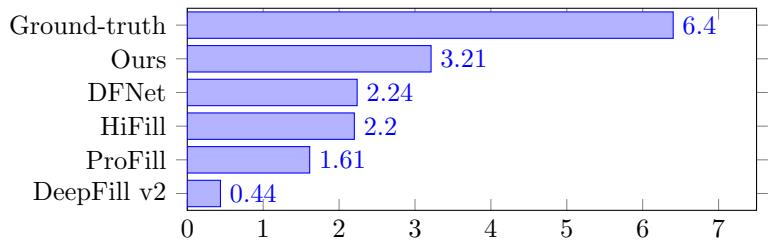

\subsubsection{Subjective evaluation}
To conduct a subjective evaluation, we selected a set of 34 natural images mostly from \cite{Comparison} with a full resolution of \begin{math}{2048\times2048}\end{math}. Participants were shown two images and asked to choose the one with the highest visual quality. We also added two validation questions to exclude unscrupulous viewers. Due to the implementation limitations of the \cite{ProFill} we made two separate comparisons at  \begin{math}{2048\times2048}\end{math} and \begin{math}{1024\times1024}\end{math} resolutions.
In first comparison 150 people participated, yielding 3,750 valid votes. In second comparison 248 people participated, yielding 6200 valid votes. Then paired estimates were converted into numerical ones using \cite{Bradley} model. The results appear in Fig.~\ref{fig41}, Fig.~\ref{fig42}.

\subsubsection{Objective evaluation}
Due to the complexity of subjective evaluation, we only conducted objective comparisons for other resolutions, although this may not be confirmed by observers ratings. For the objective comparison, we also added output images from Adobe Photoshop 2020, a commercial package that implements the classical inpainting approach. Our quality metrics were the mean L1 distance, SSIM \cite{SSIM}, and PSNR. To reduce resolution, we used Nearest-Neighbor downsampling. Table~\ref{tab1} shows the results for this objective comparison.

\begin{table}[!ht]
\centering
\caption{Objective comparison with different resolutions}\label{tab1}
\begin{tabular}{lccclccc}
\hline
\multicolumn{1}{|c|}{\backslashbox{Methods}{Metrics}}          & \multicolumn{1}{c|}{L1}             & \multicolumn{1}{c|}{PSNR}            & \multicolumn{1}{c|}{SSIM}           & \multicolumn{1}{l|}{}                           & \multicolumn{1}{c|}{L1}             & \multicolumn{1}{c|}{PSNR}            & \multicolumn{1}{c|}{SSIM}           \\ \cline{1-4} \cline{6-8} 
\multicolumn{1}{|l|}{DeepFill v2}    & \multicolumn{1}{c|}{4.981}          & \multicolumn{1}{c|}{22.389}          & \multicolumn{1}{c|}{0.938}          & \multicolumn{1}{l|}{}                           & \multicolumn{1}{c|}{5.397}          & \multicolumn{1}{c|}{21.956}          & \multicolumn{1}{c|}{0.944}          \\ \cline{1-4} \cline{6-8} 
\multicolumn{1}{|l|}{DFNet}          & \multicolumn{1}{c|}{3.592}          & \multicolumn{1}{c|}{24.910}          & \multicolumn{1}{c|}{0.946}          & \multicolumn{1}{l|}{}                           & \multicolumn{1}{c|}{4.132}          & \multicolumn{1}{c|}{23.836}          & \multicolumn{1}{c|}{0.946}          \\ \cline{1-4} \cline{6-8} 
\multicolumn{1}{|l|}{HiFill}         & \multicolumn{1}{c|}{4.422}          & \multicolumn{1}{c|}{23.667}          & \multicolumn{1}{c|}{0.935}          & \multicolumn{1}{l|}{}                           & \multicolumn{1}{c|}{4.373}          & \multicolumn{1}{c|}{23.738}          & \multicolumn{1}{c|}{0.943}          \\ \cline{1-4} \cline{6-8} 
\multicolumn{1}{|l|}{Photoshop 2020} & \multicolumn{1}{c|}{4.115}          & \multicolumn{1}{c|}{23.789}          & \multicolumn{1}{c|}{0.944}          & \multicolumn{1}{l|}{}                           & \multicolumn{1}{c|}{13.320}         & \multicolumn{1}{c|}{23.756}          & \multicolumn{1}{c|}{0.949}          \\ \cline{1-4} \cline{6-8} 
\multicolumn{1}{|l|}{ProFill}        & \multicolumn{1}{l|}{3.906}          & \multicolumn{1}{l|}{24.677}          & \multicolumn{1}{l|}{\textbf{0.950}} & \multicolumn{1}{l|}{}                           & \multicolumn{1}{c|}{—}              & \multicolumn{1}{c|}{—}               & \multicolumn{1}{c|}{—}              \\ \cline{1-4} \cline{6-8} 
\multicolumn{1}{|l|}{Ours}           & \multicolumn{1}{c|}{\textbf{3.524}} & \multicolumn{1}{c|}{\textbf{25.175}} & \multicolumn{1}{c|}{0.944}          & \multicolumn{1}{l|}{}                           & \multicolumn{1}{c|}{\textbf{3.474}} & \multicolumn{1}{c|}{\textbf{25.311}} & \multicolumn{1}{c|}{\textbf{0.950}} \\ \cline{1-4} \cline{6-8} 
\multicolumn{1}{|l|}{Resolution}     & \multicolumn{3}{c|}{1024x1024}                                                                                   & \multicolumn{1}{l|}{}                           & \multicolumn{3}{c|}{1536x1536}                                                                                   \\ \hline
\multicolumn{8}{l}{}                                                                                                                                                                                                                                                                                                         \\ \hline
\multicolumn{1}{|c|}{\backslashbox{Methods}{Metrics}}          & \multicolumn{1}{c|}{L1}             & \multicolumn{1}{c|}{PSNR}            & \multicolumn{1}{c|}{SSIM}           & \multicolumn{1}{l|}{\multirow{7}{*}{\textbf{}}} & \multicolumn{1}{c|}{L1}             & \multicolumn{1}{c|}{PSNR}            & \multicolumn{1}{c|}{SSIM}           \\ \cline{1-4} \cline{6-8} 
\multicolumn{1}{|l|}{DeepFill v2}    & \multicolumn{1}{c|}{5.546}          & \multicolumn{1}{c|}{21.834}          & \multicolumn{1}{c|}{0.946}          & \multicolumn{1}{l|}{}                           & \multicolumn{1}{c|}{5.669}          & \multicolumn{1}{c|}{21.697}          & \multicolumn{1}{c|}{0.948}          \\ \cline{1-4} \cline{6-8} 
\multicolumn{1}{|l|}{DFNet}          & \multicolumn{1}{c|}{4.345}          & \multicolumn{1}{c|}{23.424}          & \multicolumn{1}{c|}{0.948}          & \multicolumn{1}{l|}{}                           & \multicolumn{1}{c|}{4.633}          & \multicolumn{1}{c|}{22.915}          & \multicolumn{1}{c|}{0.950}          \\ \cline{1-4} \cline{6-8} 
\multicolumn{1}{|l|}{HiFill}         & \multicolumn{1}{c|}{4.403}          & \multicolumn{1}{c|}{23.696}          & \multicolumn{1}{c|}{0.944}          & \multicolumn{1}{l|}{}                           & \multicolumn{1}{c|}{4.391}          & \multicolumn{1}{c|}{23.710}          & \multicolumn{1}{c|}{0.947}          \\ \cline{1-4} \cline{6-8} 
\multicolumn{1}{|l|}{Photoshop 2020} & \multicolumn{1}{c|}{4.047}          & \multicolumn{1}{c|}{23.863}          & \multicolumn{1}{c|}{\textbf{0.952}} & \multicolumn{1}{l|}{}                           & \multicolumn{1}{c|}{4.153}          & \multicolumn{1}{c|}{23.659}          & \multicolumn{1}{c|}{0.952}          \\ \cline{1-4} \cline{6-8} 
\multicolumn{1}{|l|}{Ours}           & \multicolumn{1}{c|}{\textbf{3.447}} & \multicolumn{1}{c|}{\textbf{25.374}} & \multicolumn{1}{c|}{\textbf{0.952}} & \multicolumn{1}{l|}{}                           & \multicolumn{1}{c|}{\textbf{3.434}} & \multicolumn{1}{c|}{\textbf{25.412}} & \multicolumn{1}{c|}{\textbf{0.954}} \\ \cline{1-4} \cline{6-8} 
\multicolumn{1}{|l|}{Resolution}     & \multicolumn{3}{c|}{1792x1792}                                                                                   & \multicolumn{1}{l|}{}                           & \multicolumn{3}{c|}{2048x2048}                                                                                   \\ \hline
\end{tabular}
\end{table}

\subsection{GIMP Plugin}
Inspired by \cite{soman2020GIMPML}, we embedded our method in the GNU Image Manipulation Program (GIMP). The final plugin, which implements our method, appears in the repository.

\section{Conclusion}
This work proposes an inpainting technique that handles images of different sizes. We conducted both objective and subjective comparisons with existing learning-based models and a popular commercial package, showing that our method produces a more satisfactory result. Also, our approach can theoretically apply to any inpainting model, making it resolution independent. In the future, we would like to train a new model using the proposed method, but in an end-to-end manner with dynamic shift size.

\subsubsection {Acknowledgments}
This work was partially supported by Russian Foundation for Basic Research under Grant 19-01-00785 a. Model training for this work employed the \href{https://hpc.cs.msu.ru/polus}{IBM Polus} computing cluster of the Faculty of Computational Mathematics and Cybernetics at Moscow State University.

\begin{figure}[!ht]
\centering
\setlength\tabcolsep{1pt}
\settowidth\rotheadsize{Radcliffe Cam}
\setkeys{Gin}{width=\hsize}
\begin{tabularx}{1.0\linewidth}{l XXXX}
&   \includegraphics[valign=m]{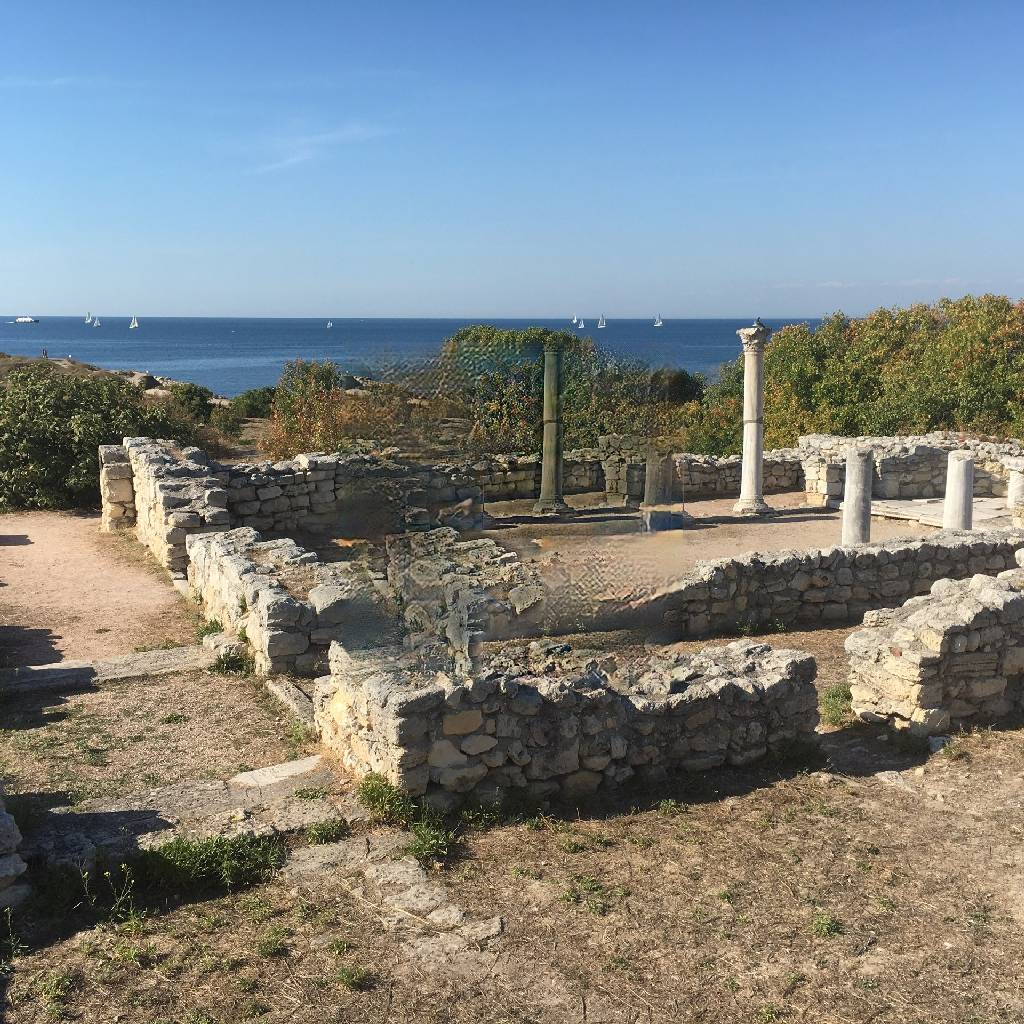}  
&   \includegraphics[valign=m]{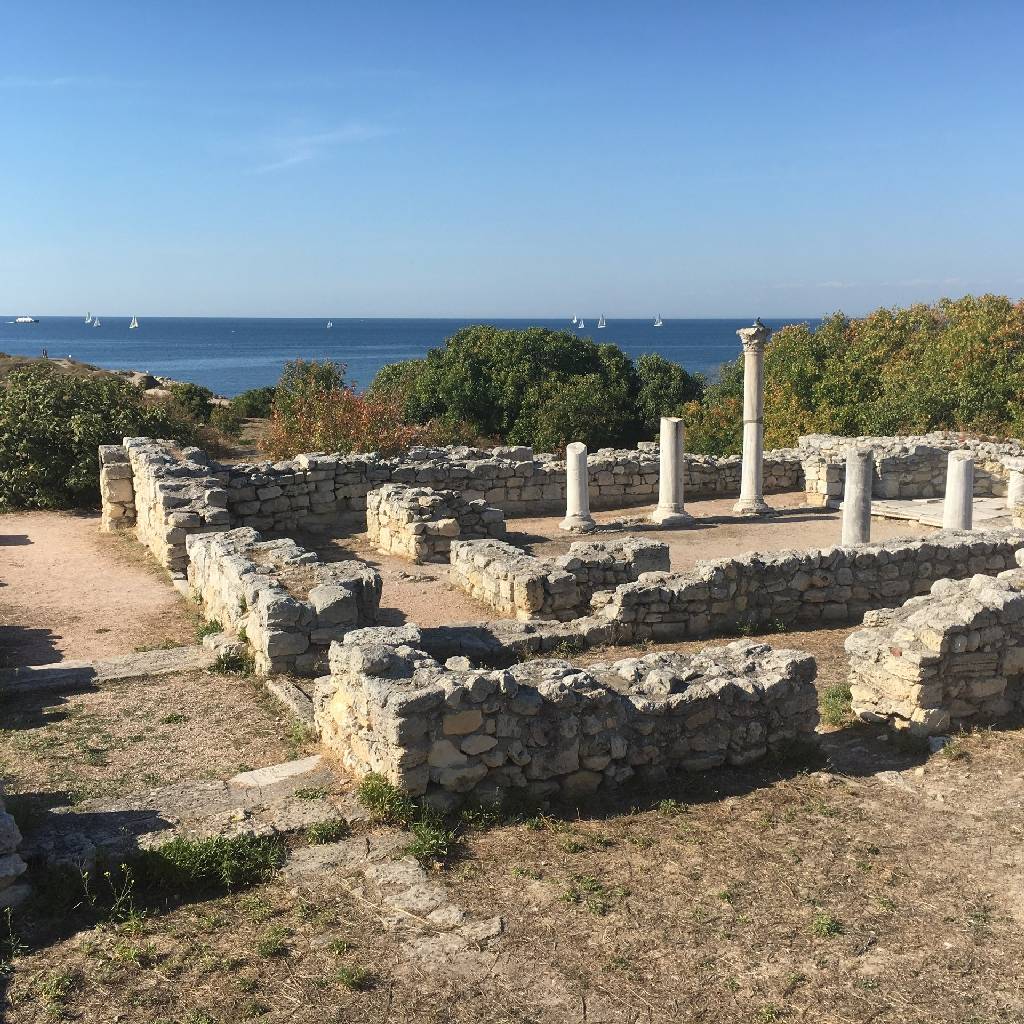}
&   \includegraphics[valign=m]{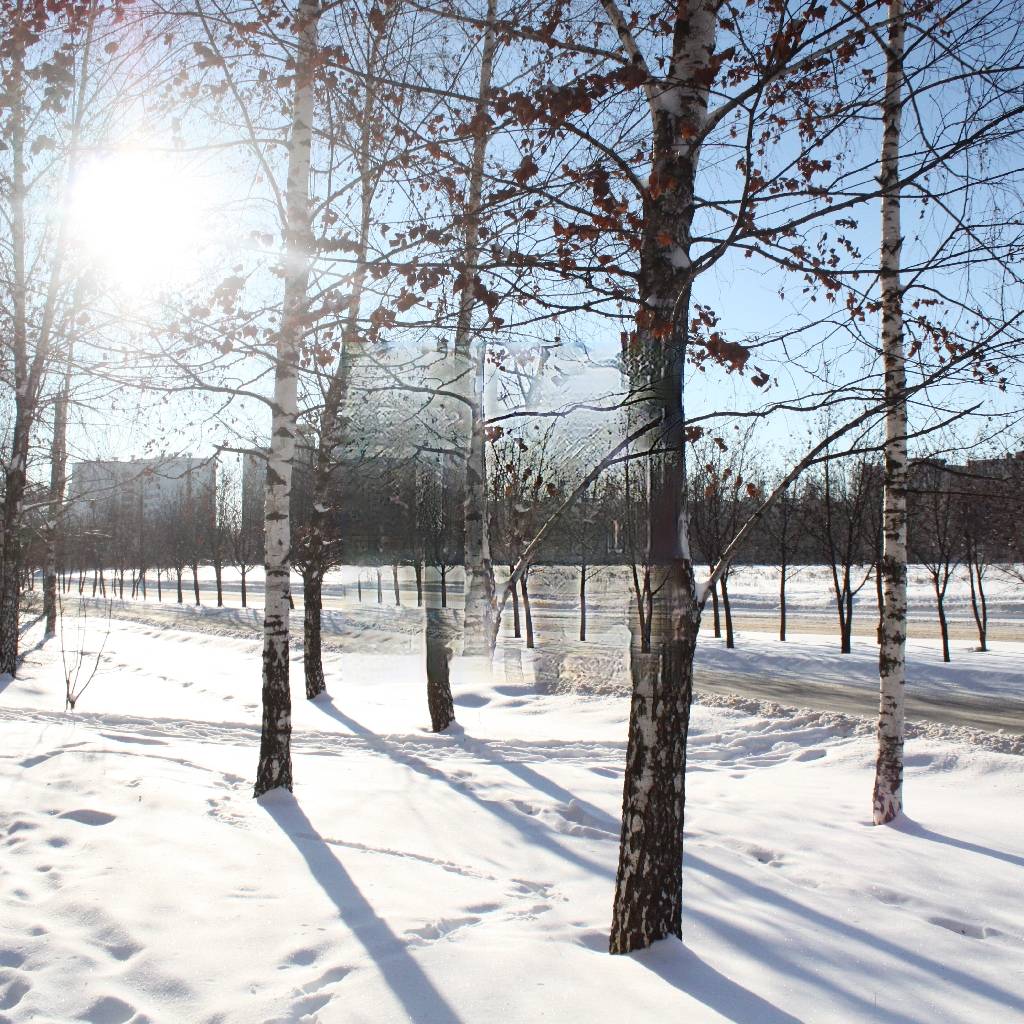}  
&   \includegraphics[valign=m]{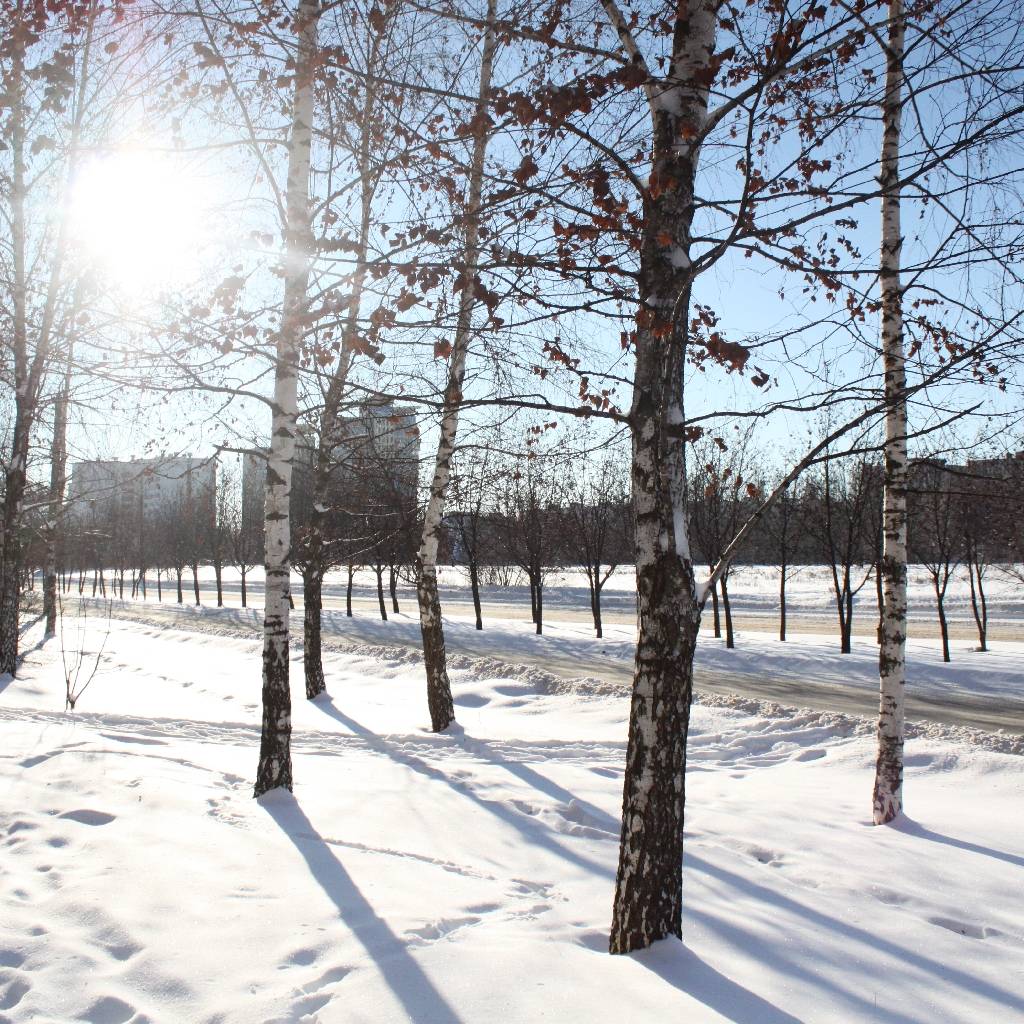}  \\
\addlinespace[2pt]
&   \includegraphics[valign=m]{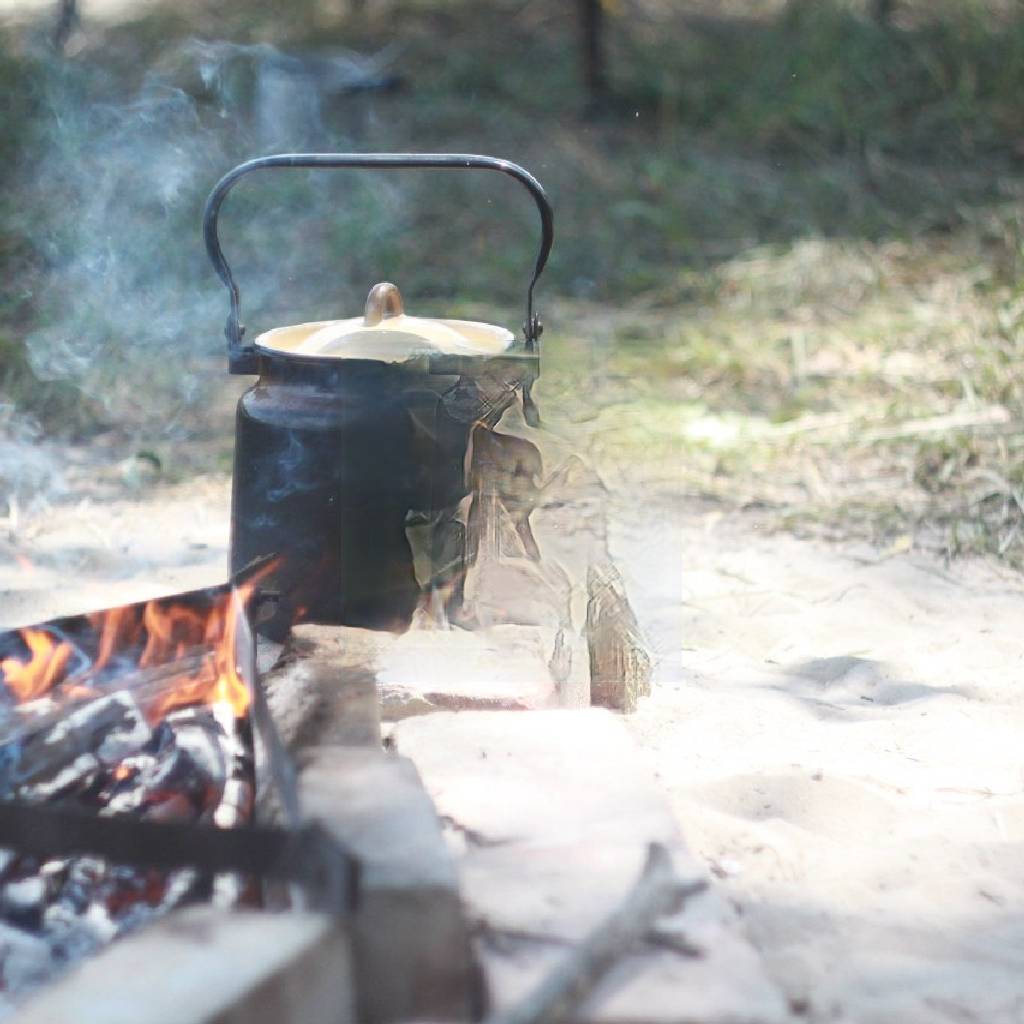}
&   \includegraphics[valign=m]{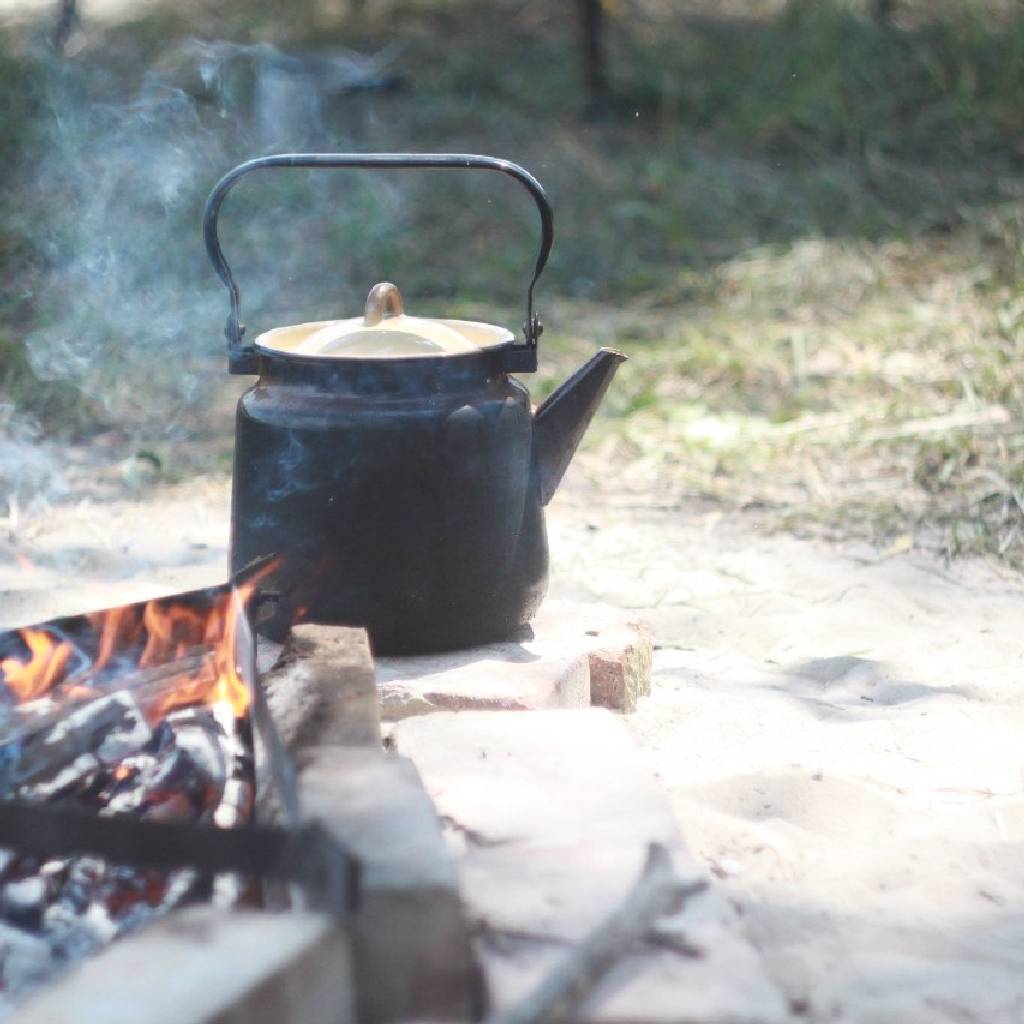}
&   \includegraphics[valign=m]{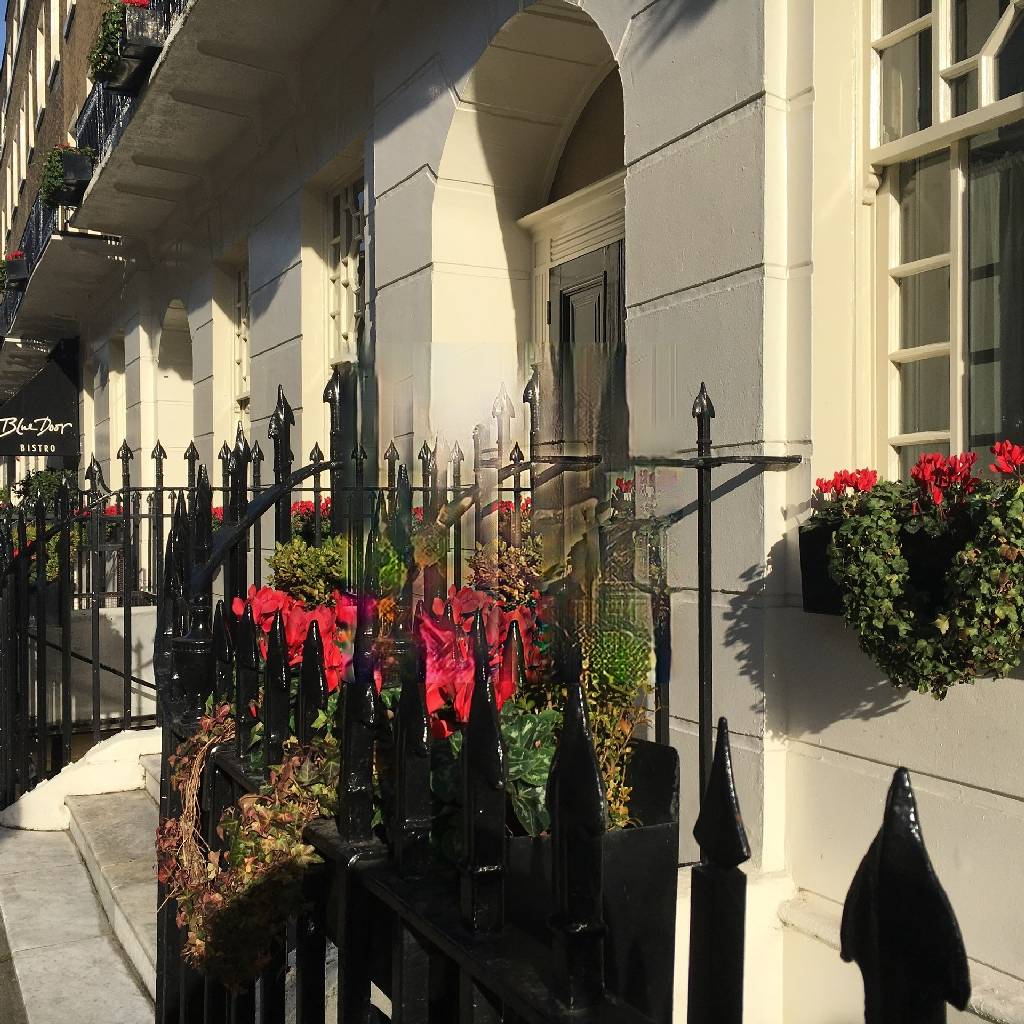}  
&   \includegraphics[valign=m]{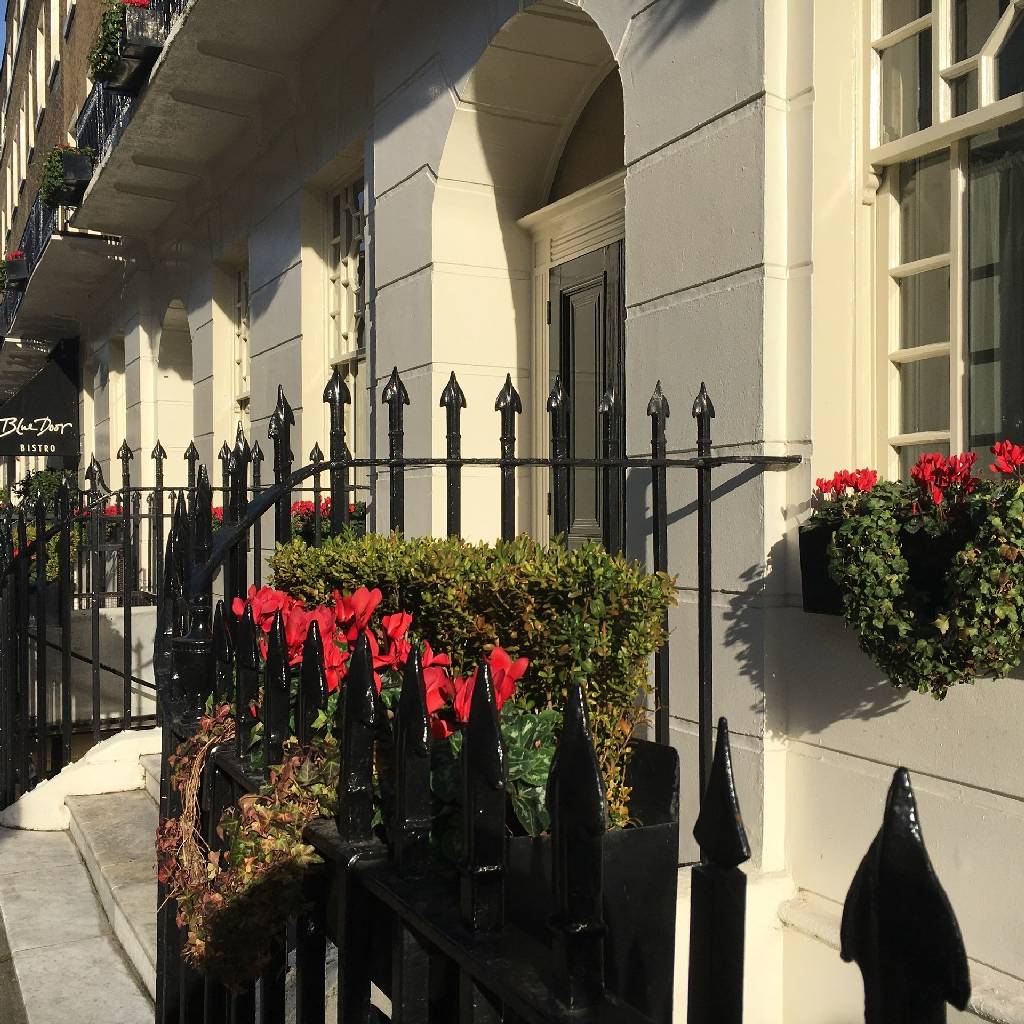}  \\
\textbf{} &
\textbf{Ours} &
\textbf{GT} &
\textbf{Ours} &
\textbf{GT}
\end{tabularx}
\caption{Failures of our model}
\label{fig6}
\end{figure}

\bibliographystyle{splncs04}
\bibliography{references}

\begin{thebibliography}{10}
\providecommand{\url}[1]{\texttt{#1}}
\providecommand{\urlprefix}{URL }

\bibitem{FragmentBased}
Drori, I., Cohen-Or, D., Yeshurun, H.: Fragment-based image completion. ACM
  Transactions on Graphics  22 (08 2003)

\bibitem{Criminisi}
{Criminisi}, A., {Perez}, P., {Toyama}, K.: Region filling and object removal
  by exemplar-based image inpainting. IEEE Transactions on Image Processing
  13(9),  1200--1212 (2004)

\bibitem{PatchMatch}
Barnes, C., Shechtman, E., Finkelstein, A., Goldman, D.: Patchmatch: A
  randomized correspondence algorithm for structural image editing. ACM Trans.
  Graph.  28 (08 2009)

\bibitem{Irregular}
Liu, G., Reda, F.A., Shih, K.J., Wang, T.C., Tao, A., Catanzaro, B.: Image
  inpainting for irregular holes using partial convolutions. In: The European
  Conference on Computer Vision (ECCV) (2018)

\bibitem{Genv1}
Yu, J., Lin, Z., Yang, J., Shen, X., Lu, X., Huang, T.S.: Generative image
  inpainting with contextual attention. 2018 IEEE/CVF Conference on Computer
  Vision and Pattern Recognition  (Jun 2018)

\bibitem{DFNet}
Hong, X., Xiong, P., Ji, R., Fan, H.: Deep fusion network for image completion.
  In: Proceedings of the 27th ACM International Conference on Multimedia. pp.
  2033--2042. MM '19, ACM, New York, NY, USA (2019)

\bibitem{Comparison}
Molodetskikh, I., Erofeev, M., Vatolin, D.: Perceptually motivated method for
  image inpainting comparison (2019)

\bibitem{GANs}
Goodfellow, I., Pouget-Abadie, J., Mirza, M., Xu, B., Warde-Farley, D., Ozair,
  S., Courville, A., Bengio, Y.: Generative adversarial nets. ArXiv  (06 2014)

\bibitem{globally}
Iizuka, S., Simo-Serra, E., Ishikawa, H.: Globally and locally consistent image
  completion. ACM Transactions on Graphics  36,  1--14 (07 2017)

\bibitem{Genv2}
Yu, J., Lin, Z., Yang, J., Shen, X., Lu, X., Huang, T.: Free-form image
  inpainting with gated convolution. 2019 IEEE/CVF International Conference on
  Computer Vision (ICCV)  (Oct 2019)

\bibitem{HiFill}
Yi, Z., Tang, Q., Azizi, S., Jang, D., Xu, Z.: Contextual residual aggregation
  for ultra high-resolution image inpainting (2020)

\bibitem{ProFill}
Zeng, Y., Lin, Z., Yang, J., Zhang, J., Shechtman, E., Lu, H.: High-resolution
  image inpainting with iterative confidence feedback and guided upsampling.
  In: European Conference on Computer Vision. pp. 1--17. Springer (2020)

\bibitem{Ronneberger_2015}
Ronneberger, O., Fischer, P., Brox, T.: U-net: Convolutional networks for
  biomedical image segmentation. Medical Image Computing and Computer-Assisted
  Intervention – MICCAI 2015 p. 234–241 (2015)

\bibitem{VGG16}
Simonyan, K., Zisserman, A.: Very deep convolutional networks for large-scale
  image recognition. arXiv 1409.1556  (09 2014)

\bibitem{DIV2K}
Timofte, R., Gu, S., Wu, J., Van~Gool, L., Zhang, L., Yang, M.H., Haris, M.,
  et~al.: Ntire 2018 challenge on single image super-resolution: Methods and
  results. In: The IEEE Conference on Computer Vision and Pattern Recognition
  (CVPR) Workshops (June 2018)

\bibitem{ioffe2015batch}
Ioffe, S., Szegedy, C.: Batch normalization: Accelerating deep network training
  by reducing internal covariate shift (2015)

\bibitem{Johnson_2016}
Johnson, J., Alahi, A., Fei-Fei, L.: Perceptual losses for real-time style
  transfer and super-resolution. Lecture Notes in Computer Science p. 694–711
  (2016)

\bibitem{kingma2014adam}
Kingma, D.P., Ba, J.: Adam: A method for stochastic optimization (2014)

\bibitem{Bradley}
Bradley, R.A., Terry, M.E.: Rank analysis of incomplete block designs: I. the
  method of paired comparisons. Biometrika  39(3/4),  324--345 (1952)

\bibitem{SSIM}
Wang, Z., Bovik, A., Sheikh, H., Simoncelli, E.: Image quality assessment: From
  error visibility to structural similarity. Image Processing, IEEE
  Transactions on  13,  600 -- 612 (05 2004)

\bibitem{soman2020GIMPML}
Soman, K.: Gimp-ml: Python plugins for using computer vision models in gimp.
  arXiv preprint arXiv:2004.13060  (2020)

\end{thebibliography}

\end{document}